\documentclass[pra,aps,twocolumn,nofootinbib,groupedaddress]{revtex4-2} %superscriptaddress

\usepackage{graphicx}
\usepackage[nointegrals]{wasysym} %
\usepackage[export]{adjustbox}
\usepackage{amsmath,amsfonts,amssymb,latexsym}
\usepackage{hhline}
\usepackage{bm}
\usepackage{verbatim}
\usepackage{enumitem}
\usepackage{mathrsfs}
\usepackage{slashed}
\usepackage{empheq}
\usepackage{physics}

\usepackage{graphicx}
\usepackage{dcolumn}
\usepackage{bm}
\usepackage{multirow}

\usepackage[normalem]{ulem}

\usepackage{amsmath}
\usepackage{amsthm}
\usepackage{amstext}
\usepackage{amssymb}
\usepackage{mathrsfs}
\usepackage{amsfonts}
\usepackage{amsbsy} 

\usepackage{braket}

\usepackage[all,matrix,cmtip]{xy}
\usepackage{mathtools}

\usepackage{color}
\definecolor{red}{rgb}{1,0,0}
\definecolor{blue}{rgb}{0,0,1}
\definecolor{dblue}{rgb}{0,0,0.4}
\definecolor{green}{rgb}{0,1,0}
\definecolor{black}{rgb}{0,0,0}
\definecolor{white}{rgb}{1,1,1}
\definecolor{pastelblue}{RGB}{20,93,160}

\definecolor{brn}{rgb}{.8,.4,.0}
\definecolor{redo}{rgb}{1,.5,.0}
\definecolor{ddgrn}{rgb}{0,0.4,0}
\definecolor{dgrn}{rgb}{0,0.55,0}
\definecolor{dbl}{rgb}{0,0,0.5}

\usepackage[colorlinks,citecolor=pastelblue,linkcolor=pastelblue,urlcolor=pastelblue]{hyperref}

\newcommand{\Z}{\mathbb{Z}}

\newcommand{\sgn}{{\rm sgn}}

\newcommand{\ie}{{\it i.e.~}} 
\newcommand{\eg}{{\it e.g.~}}

\newcommand{\bpm}{\begin{pmatrix}}
	\newcommand{\epm}{\end{pmatrix}}
\newcommand{\bmm}{\begin{matrix}}
	\newcommand{\emm}{\end{matrix}}
\newcommand{\bvm}{\begin{vmatrix}}
	\newcommand{\evm}{\end{vmatrix}}

\newcommand{\cA}{ {\cal A} }

	\newcommand{\cM}{ {\cal M} }

\usepackage{euscript}

\newcommand{\al}{\alpha}

	\renewcommand{\th}{\theta}

	\newcommand{\vphi}{\varphi}

\makeatletter
\newsavebox{\@brx}
\newcommand{\llangle}[1][]{\savebox{\@brx}{\(\m@th{#1\langle}\)}%
	\mathopen{\copy\@brx\kern-0.5\wd\@brx\usebox{\@brx}}}
\newcommand{\rrangle}[1][]{\savebox{\@brx}{\(\m@th{#1\rangle}\)}%
	\mathclose{\copy\@brx\kern-0.5\wd\@brx\usebox{\@brx}}}
\makeatother

\allowdisplaybreaks

\usepackage{tikz}
\definecolor{myRed}{RGB}{188,0,4} % #BC0004
\definecolor{myGray}{RGB}{146,146,146} % #929292
\definecolor{myBlue}{RGB}{0,0,133} % #000085

\newcommand{\Av}{
    \resizebox{0.1\textwidth}{!}{
    \begin{tikzpicture}
    \draw[line width=2pt, myGray] (-2, 0) -- (2, 0); % horizontal axis
    \draw[line width=2pt, myGray] (0, -2) -- (0, 2); % vertical axis
    
    \tikzstyle{operator} = [scale=3, myRed]
    
    \node[operator] at (0.2125, 1) {$X^\dagger$}; % top
    \node[operator] at (1, 0.0625) {$X^\dagger$}; % right
    \node[operator] at (0, -1) {$X$}; % bottom
    \node[operator] at (-1,0) {$X$}; % left
    \end{tikzpicture}
    }% End of resizebox
}
\newcommand{\Bp}{
    \resizebox{0.1\textwidth}{!}{
    \begin{tikzpicture}
    \draw[line width=1pt, myGray] (-0.5175, 0.5175) -- (-0.5175, -0.5175); % left
    \draw[line width=1pt, myGray] (0.5175, 0.5175) -- (0.5175, -0.5175); % right
    \draw[line width=1pt, myGray] (-0.5, -0.5) -- (0.5, -0.5); % bottom
    \draw[line width=1pt, myGray] (-0.5, 0.5) -- (0.5, 0.5); % top
    
    \tikzstyle{operator} = [scale=1.5, myBlue]
    
    \node[operator] at (0, 0.5) {$Z$}; % top
    \node[operator] at (0.6, 0.0425) {$Z^\dagger$}; % right
    \node[operator] at (0.11, -0.5) {$Z^\dagger$}; % bottom
    \node[operator] at (-0.5,0) {$Z$}; % left
    \end{tikzpicture}
    }% End of resizebox
}

\begin{document}

%\preprint{APS/123-QED}

\title{Analytic framework for self-dual criticality in \texorpdfstring{$\mathbb{Z}_k$}{Zk} gauge theory with matter}% Force line breaks with \\

\author{Zhengyan Darius Shi}%
\email{zdshi@mit.edu}
\affiliation{
Department of Physics, Massachusetts Institute of Technology,
Cambridge, Massachusetts 02139, USA
}

\author{Arkya Chatterjee}
\email{achatt@mit.edu}
 %\homepage{}
\affiliation{
Department of Physics, Massachusetts Institute of Technology,
Cambridge, Massachusetts 02139, USA
}%

\date{\today}

\begin{abstract}
    The deconfined phase of 2+1D $\mathbb{Z}_k$ gauge theory exhibits topological order, with $e$ and $m$ anyons that have a $2\pi/k$ braiding phase.
    Proliferating either $e$ or $m$ drives Higgs or confinement transitions, respectively.
    At the multicritical point where these transitions meet, the theory enjoys an additional duality symmetry that exchanges $e$ and $m$ anyons. 
    This symmetry forces anyons with nontrivial braiding to close their gaps simultaneously, giving rise to a critical theory that mixes strong interactions with mutual statistics. 
    We propose an effective U(1) $\times$ U(1) gauge theory with a mutual Chern-Simons term at level $k$ to describe the vicinity of the multicritical point for $k \geq 4$. The emergence of a global U(1)$^{\rm{top}}$  $\times$ U(1)$^{\rm{top}}$ symmetry at the critical point imposes powerful constraints on universal properties of the phase transition. In particular, we show that (1) the lattice magnetic flux operator embeds as a conserved U(1) current with protected scaling dimension; (2) the first-order line emanating from the critical point for $k = 2$ disappears generically for sufficiently large $k$; (3) the correlation length exponent approaches that of the 3D XY model with corrections of order $1/k^2$ in the large $k$ limit. These predictions can be tested in near-term numerical simulations and pave the way for a more general exploration of topological quantum criticality enriched with anyon-permuting symmetries.
\end{abstract}

%\keywords{Suggested keywords}%Use showkeys class option if keyword display desired
\maketitle

\paragraph*{Introduction:}

The classification and characterization of continuous quantum phase transitions is one of the central goals in equilibrium many-body physics. In recent decades, an increasing number of phase transitions have been discovered that challenge the conventional Landau classification based on symmetry-breaking (see~\cite{S202312638} and references therein). A particularly interesting example is the transition between phases with the same symmetry-breaking pattern but distinct topological orders~\cite{WW1993,CFW1993,SMFc9902062,RGc9906453,Wc9908394}. 
Conceptually, topological transitions have the potential to probe the interplay between nontrivial braiding statistics and strong interactions, a special feature that is not shared by other non-Landau critical points. Moreover, in light of the recent discovery of fractional quantum anomalous Hall phases in tunable 2D materials~\cite{PCAZZLWHHLTWCCFYCCXX230802657,CAWZLHZFTWRCFXYX230408470,XSJLXLGWTTJSJZLL230806177,ZXKZKVWTMS230500973,LHYRYSWTFJ230917436}, some classes of topological transitions may even be experimentally accessible in the near future, providing further motivation for a systematic understanding of their theoretical structure.

The simplest scenario for topological quantum criticality involves the condensation of an Abelian order-$n$ boson (see~\cite{B170604940} and references therein). These transitions belong to the same universality class as that of an $n$-state Potts/clock model~\cite{BS08080627,BSS11041701}, as exemplified \eg by the Higgs and confinement transitions of $\Z_k$ gauge theory. Much more mysterious is the situation where the microscopic Hamiltonian enjoys an additional anyon-permuting symmetry (APS) that exchanges condensable anyons with mutual statistics~\cite{B10041838,KTH14036478,THF150306812,TLF150606754}. Imposing such a symmetry forces the gaps of multiple anyons to close simultaneously despite their non-trivial braiding, leading to a distinct universality class. 

In this letter, we develop a theoretical framework for a class of APS-enriched topological phase transitions: 
$\Z_k$ gauge theory with scalar matter enriched by a $\Z_2$ duality symmetry that exchanges the $e$ and $m$ anyons. 
Away from the self-dual line, the transition out of the deconfined phase belongs to the $\Z_k^*$ universality class~\cite{FS1979,SSNHS0301297,XPK240200127,AAH240514830}.\footnote{
By $\Z_k^*$, we mean the $\Z_k$ symmetry-breaking transition with all operators except $\Z_k$ singlets projected out. At a more abstract level, to go from the $\Z_k$ universality class to the $\Z_k^*$ one, we gauge the $\Z_k$ symmetry of the former (see \eg \cite{SSN240304025} for a recent exposition on this terminology).
The transition for ${k = 3}$ is known to be first-order (see e.g.~\cite{SDOVS11103632}) and will be not considered here.} 
However, the transition that preserves self-duality is more challenging. Although a series of recent numerical works have established the continuity of the self-dual transition for ${k = 2}$ and the emergence of conformal invariance~\cite{TKPS08043175,SSN201215845,BPV211201824,OKGR231117994}, an analytical understanding is lacking (see however~\cite{VDS08070487,DKOSV10121740} for high-order series expansion results that access certain critical exponents). 

Here, we propose a continuum field theory description of the self-dual critical point and its neighboring phases at all ${k \geq 4}$. Using the field theory, we provide compelling evidence that the first-order line emanating from the self-dual critical point for ${k = 2}$ disappears when ${k \geq 4}$. Moreover, a U(1)$^{\rm{top}}$ $\times$ U(1)$^{\rm{top}}$ global symmetry emerges precisely at the critical point and constrains various universal properties of the phase transition. In particular, the lattice magnetic flux operator in the $\mathbb{Z}_k$ gauge theory embeds into the continuum field theory as a conserved current of the emergent symmetry with protected scaling dimension. This symmetry also facilitates a perturbative calculation of the correlation length exponent at the phase transition. Remarkably, this exponent converges to that of the 3D XY model in the large $k$ limit, with corrections that are suppressed by $1/k^2$. 

Overall, our field-theoretic analysis provides a number of sharp predictions about the self-dual critical point which are testable in near-term numerics. In what follows, we explain the arguments leading to these conclusions.

\begin{figure}
    \centering
    \includegraphics[width=0.7\linewidth]{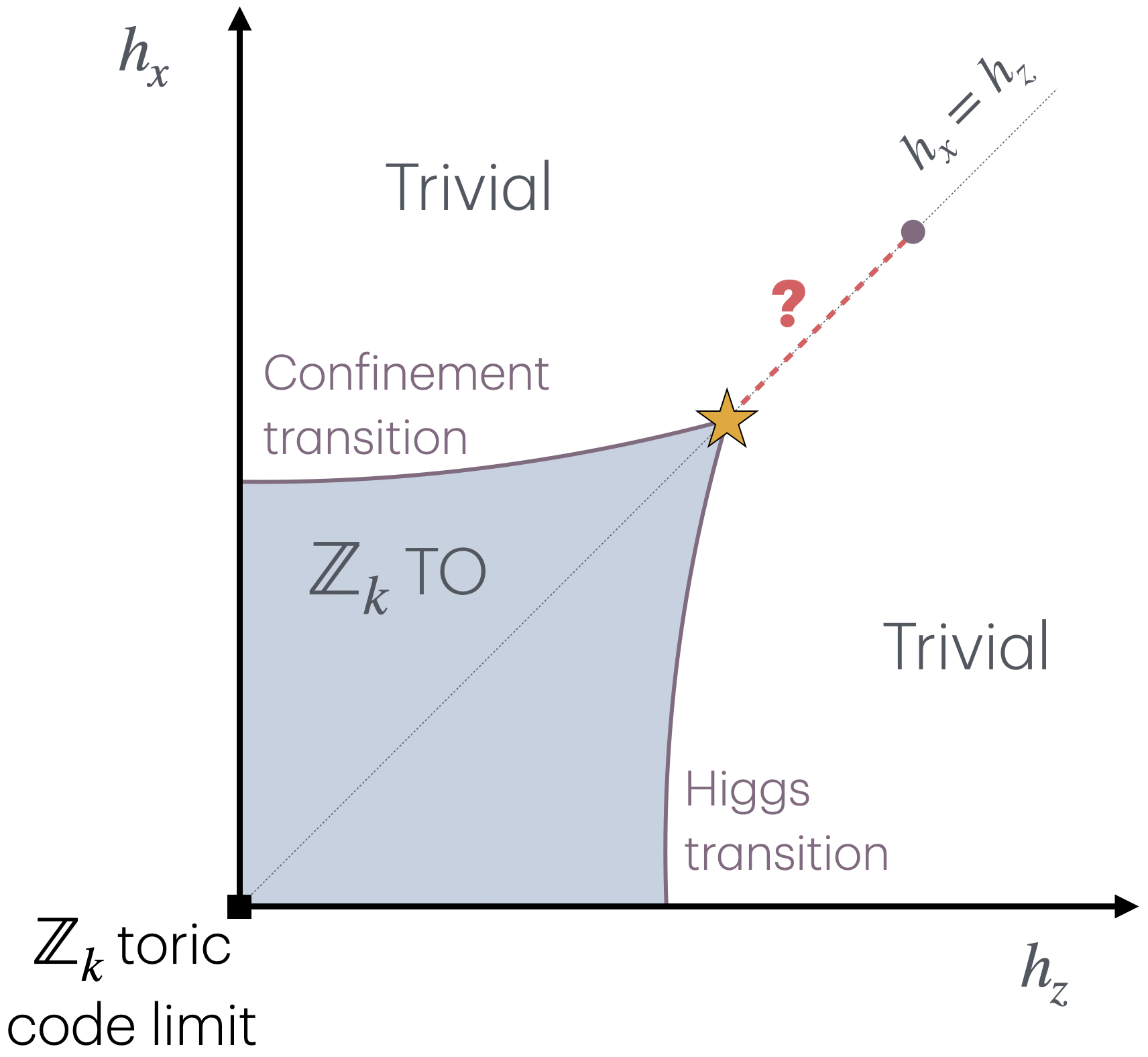}
    \caption{A proposed schematic phase diagram of $\Z_k$ gauge theory coupled to matter. The $\Z_k$ toric code limit is realized at $(h_x,h_z)=(0,0)$. The self-dual multicritical point is labeled with a star. A first-order transition on the self-dual line is found in the case of ${k=2}$, but may not exist for $k\geq 4$.}
    \label{fig:Zk-gt-pd}
\end{figure}

\paragraph*{Phase diagram and an effective Lagrangian:}
We begin with a brief review of $\Z_k$ gauge theory coupled to matter carrying the fundamental representation, which realizes $\Z_k$ topological order (TO) in its deconfined phase. A simple lattice construction of the theory---a deformation of the $\Z_k$ toric code---arises from gauging a $\Z_k$ clock model on the square lattice and fixing to unitary gauge,\footnote{For $k = 4$, there is an exact mapping between the standard $\Z_4$ clock model and two copies of Ising models, which changes the phase diagram upon gauging~\cite{S1967,GLN1981}. To avoid this fine-tuned case, the $\Z_4$ symmetric Hamiltonian prior to gauging should be perturbed away from the $\Z_k$ clock model.} 
\begin{equation}\label{eq:Zk gt}
    \mathcal{H}_{k} = 
    -\sum_{v\in V} 
    A_v
    -\sum_{p\in P} 
    B_p
    - \sum_{\ell\in L} \left(h_x \, X_\ell +h_z \,  Z_\ell  \right) +\mathrm{H.c.}
\end{equation}
where $V,P,L$ are the sets of all vertices, plaquettes, and links, $Z,X$ are $\mathbb{Z}_k$ clock and shift operators, and
\begin{equation}
     A_v = \raisebox{-0.45\height}{\Av }
    \ , 
    \qquad 
     B_p = 
     \hspace*{-5pt}
     \raisebox{-0.45\height}{\Bp }
    .
\end{equation} 
In \eqref{eq:Zk gt}, the $h_x$ and $h_z$ couplings capture the strength of gauge and matter fluctuations. In the 2-dimensional $h_x$-$h_z$ phase diagram (see Fig.~\ref{fig:Zk-gt-pd}), the region with $h_x,h_z\ll 1$ realizes the deconfined phase with $\Z_k$ TO. The $k^2$ anyons of this TO are generated by order-$k$ anyons, $e$ and $m$, which have trivial self-statistics and a mutual braiding phase of $e^{i\th_{e,m}} = e^{2 \pi i/k}$.
The general theory of topological orders tells us that the condensation of a \emph{Lagrangian condensable algebra} (LCA) trivializes the anyon content~\cite{KK11045047,K13078244}. For Abelian TOs, an LCA is essentially a composite anyon $\cA = \oplus_{a}\,  n_a \, a $,
where $n_{\mathbf 1}=1$ and the simple anyons $a$ with $n_a \neq 0$ must have trivial self and mutual statistics.
For $\Z_k$ TO, two of the LCAs are the so-called electric and magnetic LCAs,
\begin{equation}
    \cA_e = \oplus_{j=0}^{k-1} \ e^j\, , 
    \quad 
    \cA_m = \oplus_{j=0}^{k-1} \ m^j \, , 
\end{equation}
whose condensations drive transitions out of the deconfined phase via the Higgs and the confinement transitions, respectively. 
These two condensation pathways are colloquially referred to as ``$e$ condensation" and ``$m$ condensation" respectively.
The regions in the phase diagram past the Higgs and the confinement transitions are smoothly connected, following the classic argument in Ref.~\cite{FS1979}. 

$\Z_k$ TO has an anyon permutation symmetry generated by the exchange of $e$ and $m$ -- this is commonly referred to as the $e$-$m$ 
duality symmetry. Since $\cA_e$ and $\cA_m$ are interchanged under the $e$-$m$ duality, if/when the Higgs and confinement transitions meet at a multi-critical point (MCP), they must do so on the self-dual ${h_x=h_z}$ line, where the strengths of the matter and gauge fluctuations are equal. 
This $e$-$m$ duality is also present in the lattice model \eqref{eq:Zk gt}, where it acts as
\begin{equation}
 U_{em} =  T_{\left(\frac{\hat{\bm e}^{\,}_x}{2} +\frac{\hat{\bm e}^{\,}_y}{2}\right)} \prod_{\ell \in L_\vert}  H_\ell \prod_{\ell \in L_{-}}  H_\ell^{\, \dagger} \, .
\end{equation}
Here, $ H$ is the $\Z_k$ Hadamard operator which satisfies 
$ H ( X, Z)  H^\dagger = ( Z, X^\dagger) $, 
$L_\vert$ is the set of vertical links and $L_{-}$ the set of horizontal ones, and $ T_{\vec {\bm a}}$ is an $\vec {\bm a}$-translation operator. The action of $ U_{em}$ on the Hamiltonian \eqref{eq:Zk gt} interchanges the values of $h_x$ and $h_z$, so we identify it with the continuum $e$-$m$ APS of $\Z_k$ TO.

Numerical studies \cite{JSJ1980,GGRT0209027,TKPS08043175,SSN201215845,BPV211201824,OKGR231117994} have provided evidence that for $k=2$, 
the Higgs and confinement transitions indeed meet at a MCP on the self-dual line. In this work we consider the possibility that the phase diagram for $k\geq 4$ also has this self-dual MCP. The topological order of the system is completely trivialized once the system goes through any of these transitions. We postpone a discussion of the first-order line until subsequent sections.

The primary character in our story is a (Euclidean) Lagrangian description of the phase diagram in the vicinity of the self-dual MCP:
\begin{equation}
\label{eq:critical_lagrangian}
\begin{split}
L &= \sum_{\al} \left( |(i\nabla_\mu - a_\al^\mu)\phi_\al|^2 + 
    r_\al |\phi_\al^{\,}|^2 \right) 
\\
&\
+  \sum_{\al} v_\al
|\phi_\al^{\,}|^4 
+ u\, |\phi_e|^2 |\phi_m|^2
+\dots  
% \\ &\
-\frac{ik}{2\pi}\, a_e d a_m 
\\ &\, \ 
+
\sum_{q_e,q_m,n} \left( c_{(q_e,q_m)}^{(n)} 
\, \cM_{(q_e,q_m)}^{(n)} + \mathrm{H.c.}\right) 
    \, ,
\end{split}
\end{equation}
where $\dots$ refers to higher powers of $\phi_{\al}$, and $\cM_{(q_e,q_m)}^{(n)}$ are gauge-invariant monopole annihilation operators carrying magnetic charges $q_\al \in \Z/2$ corresponding to the gauge fields $a_\al$.\footnote{This convention for monopole flux quantization is commonly used in the high energy theory literature and differs from the condensed matter convention by a factor of 2.} The monopole operators ensure that the potential U(1)$^{\rm{top}}$ 0-form global symmetries associated with the monopole number conservations for each gauge field are explicitly broken, in contrast to closely related models in the spin liquid literature~\cite{XS08111220}.

Setting ${r_e,r_m>0}$ gaps out the scalar fields which can then be integrated out. The remaining terms of the Lagrangian \eqref{eq:critical_lagrangian} include the mutual Chern-Simons term and the monopole creation and annihilation operators. Since the latter are \emph{local} gauge invariant operators, their inclusion cannot (perturbatively) change the $\Z_k$ TO encoded by the mutual Chern-Simons term.
Setting ${r_{m/e}<0}$ with ${r_{e/m}>0}$ 
drives a Higgs transition for the gauge field $a_{m/e}$. 
This kills the Chern-Simons term and the proliferation of monopoles leads to a trivial gapped phase. 
Without the inclusion of monopole terms, we would have a superfluid phase associated with the spontaneous breaking of the U(1)$^{\rm{top}}$ number conservation of the $a_{m/e}$ monopoles. At ${r_{m/e}=0}$ with ${r_{e/m}>0}$, the Chern-Simons term Higgses $a_{m/e}$ from U(1) down to $\Z_k$; as a result, these critical theories belong to the $\Z_k^*$ universality class, in agreement with the existing literature~\cite{FS1979,SSNHS0301297,AAH240514830}. Finally, setting ${r_e, r_m < 0}$ gaps out both gauge fields $a_e$, $a_m$, leading to a gapped phase with trivial TO. Therefore, we conclude that the effective Lagrangian in \eqref{eq:critical_lagrangian} correctly describes the neighborhood of the self-dual MCP.

\paragraph*{Analytic theory of self-dual criticality:}%Having shown that the effective Lagrangian in \eqref{eq:critical_lagrangian} correctly describes the neighborhood of the self-dual MCP, 
We will now focus on the phase transition with ${r_{\alpha} = 0}$, $ {v_e = v_m}$, ${c^{(n)}_{(q_e,q_m)} = c^{(n)}_{(q_m,q_e)}}$, and determine its universal low energy properties. Our analysis will proceed in three steps: (1) freeze the gauge fluctuations and show that the $\mathrm{O(2)} \times \mathrm{O(2)} \rtimes \Z_2$-symmetric matter sector in \eqref{eq:critical_lagrangian} is stable against $\Z_k \times \Z_k \rtimes \Z_2$-symmetric perturbations (allowed in the UV Hamiltonian) for all ${k \geq 4}$; (2) include gauge fluctuations and show that all self-dual monopoles in the Lagrangian are irrelevant at the MCP; (3) compute the anomalous dimensions generated by the matter-gauge interactions. The basic assumption behind this 3-step procedure is that the $\mathrm{U(1)} \times \mathrm{U(1)}$ gauge theory in \eqref{eq:critical_lagrangian} can be accessed by perturbing around the pure scalar field theory at $k = \infty$, where gauge fluctuations are suppressed. Since the anomalous dimensions induced by gauge fluctuations scale as $1/k^2$, this assumption likely holds for all $k \geq 4$ but may be questionable for ${k = 2}$. Therefore, we will carry out the 3-step procedure for $k \geq 4$ and make some conjectural remarks about ${k = 2}$ along the way. With this discussion in mind, we now explain these three steps, relegating some technical details to the appendices.

In the first step, we turn off gauge fluctuations and study the scalar field Lagrangian. In addition to the terms already present in \eqref{eq:critical_lagrangian}, the $\Z_k$ gauge structure in the UV model allows for $\Z_k \times \Z_k \rtimes \Z_2$-symmetric perturbations %$\Z_k \times \Z_k$ global symmetry in addition to the $\Z_2$ duality symmetry. At the field theory level, $\Z_k \times \Z_k$ 
where $\Z_k \times \Z_k$ acts on the scalar fields as ${\phi_{\alpha} \rightarrow e^{2\pi i n_{\alpha}/k} \phi_{\alpha}}$ and $\Z_2$ exchanges $\phi_e$ with $\phi_m$. Including all symmetry-allowed terms gives a general Lagrangian of the form
\begin{equation}
    \begin{aligned}
    L_{\rm mat} &= \sum_{\alpha} \left(|\nabla \phi_{\alpha}|^2 + r |\phi_{\alpha}|^2\right) + v \sum_{\alpha} |\phi_{\alpha}|^4 \\
    &+ u\, |\phi_{e}|^2|\phi_{m}|^2  + w \sum_{\alpha} \phi_{\alpha}^{\, k} + \mathrm{H.c.} + \ldots 
    \end{aligned}
\end{equation}
where $\ldots$ denote terms with higher powers of $\phi_{\alpha}$ that will be irrelevant in the infrared (IR) limit. Let us temporarily neglect the $w$ coupling and analyze the remaining Lagrangian which contains two competing interaction terms $u$ and $v$. %The $u$ term respects the full $O(4)$ symmetry of the quadratic Lagrangian, which is broken down to $O(2) \times O(2) \rtimes Z_2$ by the presence of the $v$ term. 
The general problem of interacting CFTs with $\mathrm{O(2)} \times \mathrm{O(2)} \rtimes \Z_2$ symmetry has been studied with high-order perturbative expansions as well as conformal bootstrap. 
The simplest stable fixed point is a decoupled pair of XY models with ${u = 0}$, ${v \neq 0}$~\cite{CPV0209580,CPV9912115}. The nature of additional stable fixed points with ${u, v \neq 0}$ is under debate. While only a single such fixed point is identified in the $\epsilon$-expansion, conformal bootstrap finds two additional kinks in the allowed parameter space~\cite{S190400017,HS210108788,KS211203919}, with predictions for critical exponents that disagree with the $\epsilon$-expansion. In this work, we take the conservative approach and assume that the matter sector is close to the stable decoupled XY fixed point. 
Relative to this fixed point, the higher-order scalar operators $|\phi_{\alpha}|^6$ are strongly irrelevant and can be neglected in the IR limit~\cite{CLLPSSV191203324}.\footnote{As we later show, anomalous dimensions induced by gauge fluctuations are $1/k^2$ suppressed. Therefore, even with gauge fluctuations included, $|\phi|^6$ likely remains irrelevant for $k \geq 4$.} Furthermore, the scaling dimension of the $w$ coupling is equivalent to the scaling dimension of a $k$-fold anisotropy term in a single XY model, which is negative for all $k \geq 4$~\cite{LSB07041472}. 
Therefore, with gauge fluctuations frozen, terms that break $\mathrm{O(2)} \times \mathrm{O(2)}$ down to $\Z_k \times \Z_k$ are irrelevant and the IR theory maps to two copies of the XY model. This theory can be consistently coupled to U(1) gauge fields, resulting in the proposed Lagrangian in \eqref{eq:critical_lagrangian}. For ${k = 2}$, the complex operators $\phi_e^2$ and $\phi_m^2$ are strongly relevant and the $\mathrm{O(2)} \times \mathrm{O(2)} \rtimes \mathbb{Z}_2$ symmetric theory breaks down. 
It becomes more convenient to describe the MCP in terms of two real scalar fields $\vphi_e, \vphi_m$ coupled to a pair of $\Z_2$ gauge fields. With the $\Z_2$ gauge fields frozen, this theory flows to the XY model with an emergent $\mathrm{O(2)}$ symmetry that rotates the vector $(\vphi_e, \vphi_m)$~\cite{LF1973,FN1974,NKF1974,KNF1974}. The effects of $\Z_2$ gauge fluctuations are difficult to control analytically, and will not be discussed further in this work.

The preceding analysis provides some clues about the existence of a line of first-order transition upon exiting the deconfined phase along the self-dual line (cf. Fig.~\ref{fig:Zk-gt-pd}). %\ie ${r_e = r_m = r < 0}$, 
Such a first-order line is associated with a spontaneously broken $e$-$m$ duality symmetry, which prevents a smooth passage from the Higgs to the confinement regime.
% This first order line 
This scenario has been numerically confirmed for ${k=2}$~\cite{TKPS08043175}, but remains unexplored for ${k\geq 4}$. In the scalar field theory for ${k = 2}$ (neglecting gauge fluctuations), the leading irrelevant operator takes the form $\tilde{w} ( \vphi_e^2 -  \vphi_m^2)^2$. When $r < 0$, the inclusion of this term with $\tilde{w} < 0$ favors a ground state with $ \vphi_e^2 \neq  \vphi_m^2$, thereby spontaneously breaking the duality symmetry. 
This argument suggests that duality SSB can occur in some open region of the ${k = 2}$ phase diagram, although the region could disappear when $\tilde{w}$ is tuned from negative to positive by additional microscopic interactions that preserve the $\Z_2$ gauge structure. The situation is drastically different for ${k \geq 4}$, where the leading irrelevant operators enter the Lagrangian as
\begin{equation}
    L = L_{\rm fixed-pt} + u\, |\phi_e|^2 |\phi_m|^2 + w \sum_{\alpha} \phi_{\alpha}^{\, k}  + \mathrm{H.c.}+ \ldots 
\end{equation}
For infinitesimal $u$ and $w$, the mean-field ground state of the above Lagrangian always preserves duality symmetry, independent of the sign of $u$ and the phase of $w$. This observation leads to the conjecture that duality SSB is \textit{generically absent for $k \geq 4$}. We caution that the qualitative picture above is heuristic as it neglects the effects of gauge fluctuations. Nevertheless, the dichotomy between the ${k = 2}$ and ${k \geq 4}$ cases may be more robust and invites more careful numerical studies in the future.%}

Starting with the decoupled XY model in the matter sector, we can gauge the global $\mathrm{U(1)} \times \mathrm{U(1)}$ symmetry and study the effects of fluctuations. 
Since $\mathrm{U(1)}$ monopoles are local operators in 2+1D, it is important to calculate their scaling dimensions and assess their effects on the monopole-free critical point. This calculation is not analytically tractable for a general interacting conformal field theory. To make progress, we pass to a large-$N$ generalization of the original Lagrangian
\begin{equation}\label{eq:largeN_compact}
    \begin{aligned}
    L &= \sum_{\alpha=e,m} \sum_{i=1}^N \left[|(i\nabla - a_{\alpha}) \phi_{i,\alpha}|^2 + r|\phi_{i,\alpha}|^2\right] \\
    &\hspace{1cm}+ \frac{v}{N} \sum_{\alpha = e, m} \left(\sum_{i=1}^N |\phi_{i, \alpha}|^2\right)^2 - \frac{ik}{2\pi}\, a_e da_m \,,
    \end{aligned}
\end{equation}
and consider the limit where ${N, k \rightarrow \infty}$ with the ratio ${\kappa = k/N}$ held fixed~\cite{MS1990,MHSF08092816,P13036125,DM13091160,CMPY151107108,CP160305582,CIMP171000654,C210207377,CDW221012370}. The most general monopole operator $\mathcal{M}^{(n)}_{(q_e, q_m)}$ in this theory carries magnetic charge $(q_e, q_m)$ relative to the gauge fields $a_e, a_m$ where ${q_e, q_m \in \mathbb{Z}/2}$. From the state-operator correspondence, the lowest scaling dimension within a fixed flux sector is given by the ground state energy of the radially quantized Hamiltonian with $4\pi q_{e/m}$ fluxes of $a_{e/m}$ threaded through the spatial sphere~\cite{BKW0206054}. As usual, it is convenient to first compute the partition function ${Z_{(q_e, q_m)}(\beta, N, \kappa) = e^{-\beta F_{(q_e,q_m)}(\beta, N, \kappa)}}$ at large $N$, where the free energy $F_{(q_e, q_m)}(\beta, N, \kappa)$ admits a series expansion in $1/N$:
\begin{equation}
     F_{(q_e,q_m)}(\beta, N, \kappa) = \sum_{m=0}^{\infty} N^{1-m} F^{(m)}_{(q_e,q_m)}(\beta, \kappa) \,.
\end{equation}
The minimal monopole scaling dimension in the $(q_e, q_m)$ sector is then given by the ground state energy 
\begin{equation}
    \Delta_{(q_e,q_m)}(N, \kappa) = \lim_{\beta \rightarrow \infty} F_{(q_e,q_m)}(\beta, N, \kappa) \,. 
\end{equation}
To argue for the irrelevance of monopoles, we can restrict to the sectors $(1/2, 0)$ and $(0, 1/2)$. Monopoles with higher charge will appear in the operator product expansion of these minimal monopoles and generally have higher scaling dimensions. Since duality exchanges $(1/2,0)$ and $(0,1/2)$, these two sectors are degenerate and it suffices to consider $(1/2, 0)$ without loss of generality. %(our method works for $\mathcal{M}_{(q,0)}$ and $\mathcal{M}_{(q,q)}$ with arbitrary $q$, which will be considered in Appendix~\ref{app:monopole_leading}). 

%\ZS{Much of the discussion in the remainder of this page can probably be moved to the appendix for brevity. Try to decide how much to keep.} 
Let us now determine the free energy function $F_{(q,0)}(\beta, N, \kappa)$ in the large $N$ limit and restrict to $q = 1/2$ in the end. As in the standard treatment of critical O(N) models, the quartic interactions proportional to $v$ can be decoupled by a pair of Hubbard-Stratanovich fields $\mu_e, \mu_m$, thereby transforming the Lagrangian to\footnote{The factor of $1/4$ is fixed by conformal invariance on $S^2 \times S^1_{\beta}$.}
\begin{equation}
    L = \sum_{\alpha, i} |(i\nabla - a_{\alpha}) \phi_{\alpha,i}|^2 + \left(\frac{1}{4} + \mu_{\alpha}\right) |\phi_{\alpha}|^2 - \frac{ik}{2\pi} a_e d a_m \,. 
\end{equation}
The transformed Lagrangian is quadratic in $\phi_{\alpha}$. Therefore, the scalar fields can be integrated out, generating an effective action for $a_{\alpha}, \mu_{\alpha}$:
\begin{equation}
    \frac{S_{\rm eff}}{N} = \sum_{\alpha} \tr \log \left[(i\nabla - a_{\alpha})^2 + \frac{1}{4} + \mu_{\alpha} \right] - \frac{i\kappa}{2\pi} \int a_e d a_m \,. 
\end{equation}
We now consider a general ansatz for $a_{\alpha}, \mu_{\alpha}$ consistent with the flux assignment:
\begin{equation}
    \bar a_{\alpha,0} = - i \lambda_{\alpha} \,, \quad d \bar a_e = q \sin \theta d \theta \wedge d \phi \,, \quad d \bar a_m = 0 \,.
\end{equation}
In Appendix~\ref{app:Mq0_leading}, we solve the saddle point equations and numerically determine the scaling dimension $\Delta^{(0)}_{(q,0)}(\kappa)$ in the $(q,0)$ sector as a function of $\kappa$.
\begin{figure}
    \centering
    \includegraphics[width = 0.48\textwidth]{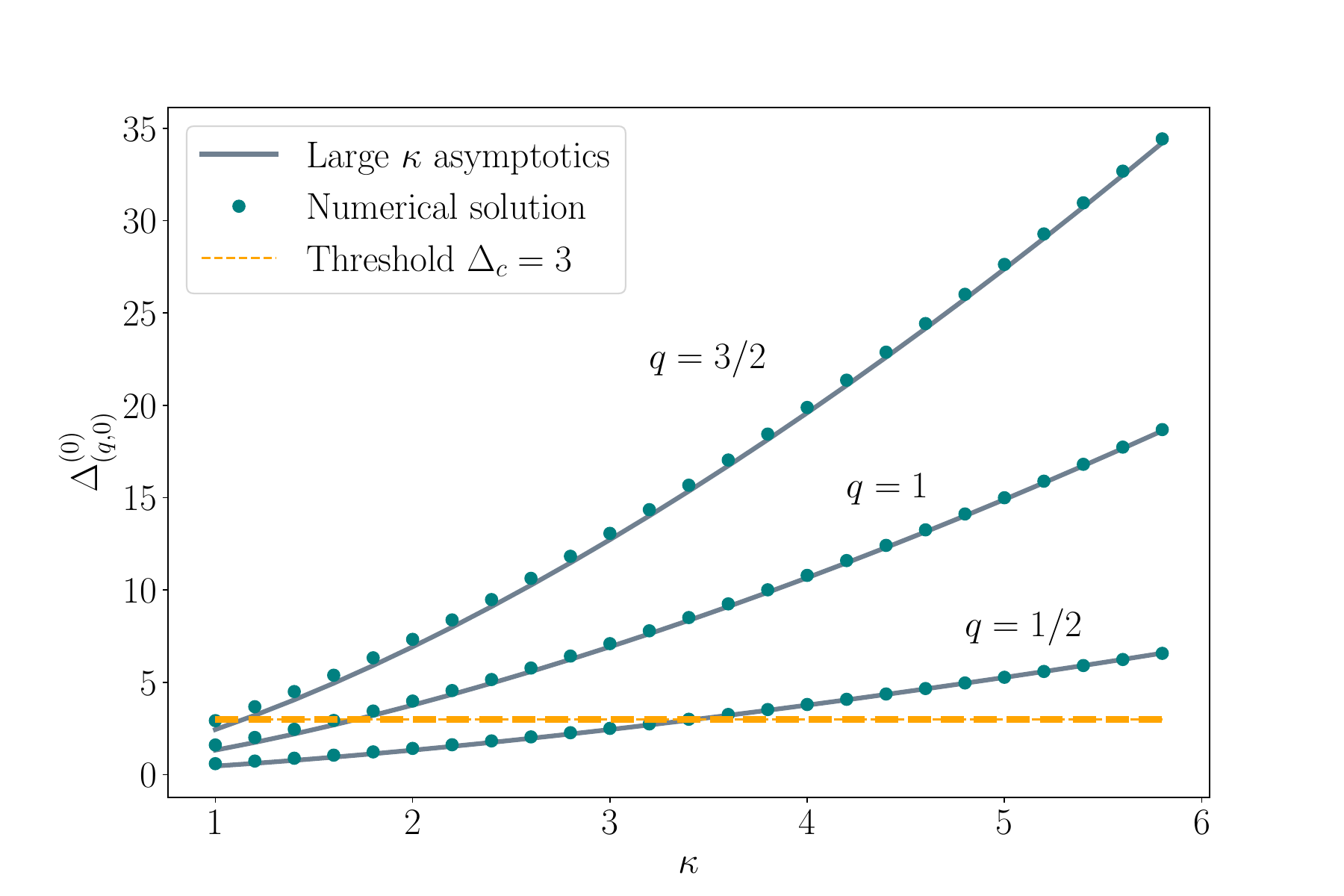}
    \caption{Saddle point solution for the minimal scaling dimension $\Delta_{(q, 0)}(N=1, \kappa)$ as a function of $\kappa = k$ in the $(q,0)$ flux sector with $q = 1/2, 1, 3/2$. Dotted curves are numerical solutions while smooth curves are analytic solutions valid in the $\kappa \rightarrow \infty$ limit.}
    \label{fig:monopole}
\end{figure}
As shown in Fig.~\ref{fig:monopole}, for ${N = 1}$, $\Delta^{(0)}_{(q,0)}(\kappa)$ exceeds the irrelevance threshold ${\Delta_c = 3}$ when ${\kappa \geq 4}$ with arbitrary $q$. Furthermore, at large $\kappa$, the numerical results agree with an asymptotically exact analytic solution (see Appendix~\ref{app:Mq0_leading})
\begin{equation}
    \mu_m(\kappa) \approx \kappa q \,, \quad \lambda_m \approx - \sqrt{\kappa q} \,, \quad \Delta^{(0)}_{(q,0)}(\kappa) \approx \frac{4}{3}(\kappa q)^{3/2} \,.
\end{equation}
Going beyond the saddle point, one can compute subleading corrections to the scaling dimensions by integrating over quadratic fluctuations of $a_{\alpha}, \mu_{\alpha}$. In Appendix~\ref{app:Mq0_subleading}, we show that in the large $\kappa$ limit, these corrections do not change the asymptotic scaling ${\Delta_{(q,0)}(\kappa) \sim \kappa^{3/2}}$, although the prefactor can be modified. The elementary monopoles $\mathcal{M}^{(n)}_{(q,0)}(\kappa)$ are therefore guaranteed to be irrelevant in the large $\kappa$ limit for any $q$. 

Composite monopoles that appear in the OPE of $\mathcal{M}^{(n_1)}_{(q_e,0)}$ and $\mathcal{M}^{(n_2)}_{(0,q_m)}$ generally have even higher scaling dimensions. As a sanity check, we analyze the lowest scaling dimension in the $(q,q)$ sector and show that it is given by a simple formula
\begin{equation}\label{eq:Deltaqq_formula}
    \Delta_{(q,q)}(N, \kappa) = 2 \Delta_{q, \rm SQED3}(N, \kappa) + \mathcal{O}(N^{-1}) \,,
\end{equation}
where $\Delta_{q, \rm SQED3}(N, \kappa)$ is the scaling dimension of a charge-$q$ monopole in scalar QED3 with a self-Chern-Simons term at level ${k = N \kappa}$ (see Appendix~\ref{app:Mqq_subleading}). Using existing results for scalar QED3 in Ref.~\cite{C210207377}, we verify that ${\Delta_{(q,q)}(N, \kappa) > \Delta_{(q,0)}(N, \kappa)}$, confirming our intuition.

The observation that monopoles are irrelevant brings us to the final step: analyze the RG flow in the monopole-free gauge theory and compute anomalous dimensions induced by the gauge-matter interaction
\begin{equation}\label{eq:non_compact_gauge_theory}
    L_{\rm int} = - \sum_{\alpha} a_{\alpha} J_{\alpha} + \sum_{\alpha} a_{\alpha}^2 |\phi_{\alpha}|^2 \,,
\end{equation}
with $J_e, J_m$ representing the $\mathrm{U(1)}$ currents associated with the matter fields $\phi_e, \phi_m$. 
To make analytic statements about this interacting CFT, we again pass to the generalized Lagrangian in \eqref{eq:largeN_compact} with $N$ flavors of $\phi_{\alpha}$ and Chern-Simons level ${k = N \kappa}$. Anomalous dimensions in this theory can be computed order by order in the $1/N$ expansion with ${\kappa = k/N}$ fixed. In Appendix~\ref{app:RG}, we carry out this computation up to $\mathcal{O}(N^{-1})$ for the leading duality symmetric/anti-symmetric operators $\sigma_{\pm}$ (which are identified with $|\phi_e|^2 \pm |\phi_m|^2$ for $k \geq 4$):
\begin{equation}\label{eq:x_AS_scaling_dimension}
    \begin{aligned}
    \Delta_{\sigma_{\pm}} &= 2 - \frac{16}{3\pi^2N} + \frac{64}{3\pi^2 N \left[1 + \frac{64 \kappa^2}{\pi^2}\right]} \\
    &\hspace{0.4cm}- \frac{64}{\pi^2 N \left[1 + \frac{64 \kappa^2}{\pi^2}\right]^2} \pm \frac{64^2 \kappa^2}{2 \pi^4 N \left[1 + \frac{64 \kappa^2}{\pi^2}\right]^2} \,.
    \end{aligned}
\end{equation}
The $\kappa \rightarrow 0$ limit recovers well-known results for the critical $\mathbb{CP}^{N-1}$ model in Refs.~\cite{HLM1974,AP1981,IKK9703011,KS08010723,BK181201544,BK190205767}. In the opposite $\kappa \rightarrow \infty$ limit, we reproduce exponents of the critical O(2N) model up to corrections that are $1/\kappa^2$ suppressed. This latter statement in fact holds in the large $k$ limit independent of $N$. The basic reason is that the gauge propagator which carries a factor of $1/k$ is off-diagonal in the $e$, $m$ indices while the interaction vertices are diagonal. This structure ensures that within a large $k$ expansion, any irreducible Feynman diagram with a single gauge propagator vanishes identically and all anomalous dimensions are at least $1/k^2$ suppressed. Importantly, the Chern-Simons level $k$ cannot be renormalized due to gauge invariance, which we verify explicitly at $\mathcal{O}(N^{-1})$ in Appendix~\ref{app:CS_level_RG}.

\paragraph*{Discussion:} In summary, we proposed an effective $\textrm{U(1)} \times \textrm{U(1)}$ gauge theory \eqref{eq:critical_lagrangian} for 2+1D $\Z_{k \geq 4}$ gauge theory with matter in the vicinity of its self-dual MCP. The $\textrm{U(1)} \times \textrm{U(1)}$ gauge structure breaks down for $k = 2$, where the existence of a useful Lagrangian description remains an important open problem. 
A striking prediction of our theory is the emergence of a global $\textrm{U(1)}^{\rm top} \times \textrm{U(1)}^{\rm top}$ symmetry at the MCP, which guarantees the existence of two current operators $j_{\alpha}^{\rm top} = \frac{1}{2\pi} \star d a_{\alpha}$ with protected scaling dimension 2.
The lattice star and plaquette operators $A_v, B_p$ have large overlap with $e^{i r^2 \nabla \times \mathbf{a}_m}, e^{i r^2 \nabla \times \mathbf{a}_e}$ (respectively) in the IR limit, where $r^2$ is the area of the lattice unit cell. Since $\nabla \times \mathbf{a}_{\alpha}$ is proportional to the temporal component of $j^{\rm top}_{\alpha}$, the emergence of a $\textrm{U(1)}^{\rm top} \times \textrm{U(1)}^{\rm top}$ symmetry can be tested by examining the decay of connected correlation functions $\ev{A_v A_{v'}}_c, \ev{B_p B_{p'}}_c$ at large spatial separation on the lattice.

The $\textrm{U(1)}^{\rm top} \times \textrm{U(1)}^{\rm top}$ symmetry also allows us to compute scaling dimensions $\Delta_{\sigma_{\pm}}$ for the most relevant duality symmetric/anti-symmetric operators $\sigma_{\pm} = |\phi_e|^2 \pm |\phi_m|^2$ and show that they approach $3 - \nu_{\rm XY}^{-1}$ as $1/k^2$ in the large $k$ limit, with $\nu_{\rm XY}$ the correlation length exponent of the 3D XY model. 
The mass terms $|\phi_e|^2$ and $|\phi_m|^2$ drive the Higgs and confinement transitions in the field theory, while the operators $Z_\ell^{\,} + Z_\ell^\dagger$ and $X_\ell^{\,} + X_\ell^\dagger$ (respectively) do so in the lattice model. So one expects $Z_\ell^{\,} + Z_\ell^\dagger$ and $X_\ell^{\,} + X_\ell^\dagger$ to have large overlap with $|\phi_e|^2$ and $|\phi_m|^2$ in the IR limit.
Following this logic, ${X_{\ell}^{\,}+ X_{\ell}^\dagger \pm (Z_{\ell}^{\,}+Z_{\ell}^\dagger)}$ should have high overlap with the duality even and odd operators $\sigma_{\pm}$.
The scaling dimensions $\Delta_{\sigma_{\pm}}$ are higher-$k$ generalizations of $x_{S/A}$ (defined in Ref.~\cite{SSN201215845}), which can be extracted from Monte Carlo numerics and compared with our analytic predictions.
Beyond Monte Carlo, it would also be interesting to realize emergent $\Z_k$ gauge theories in quantum Hall multi-layers, which can be studied using the recently proposed fuzzy sphere regularization~\cite{ZHHHH221013482}. While $\Z_k$ TO has been constructed in quantum Hall bilayers~\cite{SKWS160908616,JCX240307054,Z240312126}, a microscopic Hamiltonian that captures the entire phase diagram in Fig.~\ref{fig:Zk-gt-pd} remains a challenge.

Looking further afield, the self-dual MCP of $\Z_k$ gauge theory with matter is only the simplest realization of APS-enriched topological quantum criticality.
Generalizations of this, involving \textit{non-Abelian} boson condensation, are largely unexplored. 
For the class of TOs with a Chern-Simons theory description, some non-Abelian boson condensation transitions can be captured by a Chern-Simons-Higgs Lagrangian~\cite{CN150700344}. A wealth of analytic tools have been developed for this special class of theories~\cite{GMY11104386,AGY12074593,A151200161,ABHS161107874,SSWW160601989,SZS250502893}.
In order to study interesting APS-enriched transitions more broadly, however, it may be necessary to develop a framework that incorporates the braiding and symmetry data of the TO without requiring a field theory description.
We leave such explorations to future studies.

% \begin{acknowledgments}
\paragraph*{Acknowledgements:}
We thank Max Metlitski, Adam Nahum, Sal Pace, Akshat Pandey, T. Senthil, and Xiao-Gang Wen for helpful conversations and feedback on the draft. ZDS is supported by the Department of Energy under grant DE-SC0008739. AC is supported by NSF DMR-2022428 and by the Simons 
Collaboration on Ultra-Quantum Matter, which is a grant from the Simons 
Foundation (651446).
% \end{acknowledgments}
\bibliography{selfdual}% Produces the bibliography via BibTeX.

\onecolumngrid
\appendix

\section{Calculation of monopole scaling dimensions}

In this appendix, we provide some details on the computation of monopole scaling dimensions at the self-dual critical point. We will consider both the unbalanced monopoles labeled by $\mathcal{M}_{(q,0)}$ and the self-dual monopoles labeled by $\mathcal{M}_{(q,q)}$. The upshot is that the scaling dimensions of the elementary monopole $\mathcal{M}_{(1/2,0)}$ is already higher than the critical threshold $\Delta_c = 3$ for all $k \geq 4$. Since a general composite monopole $\mathcal{M}_{(q_e, q_m)}$ appears in the operator product expansion of $\mathcal{M}_{(q_e,0)} \times \mathcal{M}_{(0,q_m)}$ and will likely have a higher scaling dimension, this result provides strong evidence that all monopoles allowed in the self-dual Lagrangian are in fact irrelevant for all $k \geq 4$. 

Before going into the detailed computations, let us first write down the action for the self-dual theory at arbitrary $N, k$, quantized on a spatial sphere $S^2$:
\begin{equation}
    \begin{aligned}
    S &= \int d^3 x \sqrt{g} \sum_{\alpha=e,m} \sum_{i=1}^N |(i\nabla - a_{\alpha}) \phi_{i,\alpha}|^2 + (\frac{1}{4} + \mu_{\alpha}) |\phi_{i,\alpha}|^2 - \frac{ik}{2\pi} \int d^3 x a_e d a_m \,. 
    \end{aligned}
\end{equation}
Since the action is quadratic in the matter field $\phi$, we can integrate out $\phi$ to get an effective action for $a_e, a_m$ and the Hubbard-Stratanovich fields $\mu_e, \mu_m$:
\begin{equation}
    S_{\rm eff} = N \left\{\sum_{\alpha} \tr \log \left[(i\nabla - a_{\alpha})^2 + \frac{1}{4} + \mu_{\alpha}\right] - \frac{i\kappa}{2\pi}
 \int d^3 x a_e d a_m \right\} \,,
\end{equation}
Now to compute the scaling dimension of a general monopole with charge $(q_e, q_m)$ where $q_e, q_m \in \mathbb{Z}/2$, we thread $4\pi q_e$ flux of $a_e$ and $4\pi q_m$ flux of $a_m$ through the spatial $S^2$. This flux configuration defines a static background on top of which the dynamical gauge fields can fluctuate. By evaluating this path integral in a large $N$ expansion, we obtain the thermal partition function
\begin{equation}
    Z_{(q_e, q_m)}(\beta, N, \kappa = k/N) = e^{-\beta F_{(q_e,q_m)}(\beta, N, \kappa)} \,, \quad F_{(q_e, q_m)}(\beta, N, \kappa) = N F^{(0)}_{(q_e, q_m)}(\beta, \kappa) + F^{(1)}_{(q_e, q_m)}(\beta, \kappa) + \mathcal{O}(N^{-1}) \,. 
\end{equation}
In general, $Z_{(q_e, q_m)}$ does not resolve the individual energy eigenvalues, which correspond to scaling dimensions of different monopoles in the same flux sector $(q_e, q_m)$. However, for the purposes of this work, we will only be interested in the ground state (\ie the operator with the lowest scaling dimension). The ground state energy can be extracted by taking the $\beta \rightarrow \infty$ limit of the free energy:
\begin{equation}
    \Delta_{(q_e,q_m)}(N, \kappa) = \lim_{\beta \rightarrow \infty} F_{(q_e, q_m)}(\beta, N, \kappa) \,. 
\end{equation}
In the next two sections, we consider the leading order free energy $F^{(0)}_{(q_e,q_m)}$ and the subleading corrections $F^{(1)}_{(q_e,q_m)}$ in order. As we will see, $F^{(0)}_{(q_e,q_m)}$ can be evaluated efficiently for arbitrary $\kappa$, while $F^{(1)}_{(q_e,q_m)}$ requires a more complicated numerical summation. In the special case where $q_e = q_m = q$, we derive an exact relationship between $F^{(1)}_{(q, q)}$ and the corresponding subleading scaling dimension of a charge-$q$ monopole in SQED3 with a self Chern-Simons term, which has been evaluated before. Numerical solution of the more general case is left for future work. 

\subsection{Leading order: saddle point analysis}\label{app:monopole_leading}

The leading order free energy $F^{(0)}_{(q_e,q_m)}$ can be determined by solving the saddle point equations in the flux sector $(q_e, q_m)$. The most general static saddle point ansatz consistent with this flux configuration can be written as
\begin{equation}\label{eq:saddle_general_ansatz}
    \bar a_{\alpha,0}(\theta, \phi) = -i \lambda_{\alpha} \,, \quad d \bar a_{\alpha}(\theta, \phi) = q_{\alpha} \sin \theta d \theta \wedge d \phi \,, \quad \bar \mu_{\alpha}(\theta,\phi) = \mu_{\alpha}  \,.
\end{equation}
In this static background, the mutual Chern-Simons term can be evaluated as
\begin{equation}
    \begin{aligned}
    \frac{ik}{2\pi} \int_{S^2 \times S^1_{\beta}} a_e d a_m &= \frac{ik}{2\pi} \int_{S^2 \times D^2_{\beta}} f_e \wedge f_m = \frac{ik}{2\pi} (\int_{S^2} f_e) (\int_{D^2_{\beta}} f_m) + \frac{ik}{2\pi} (\int_{S^2} f_m) (\int_{D^2_{\beta}} f_e) = 2 k \beta (q_e \lambda_m + q_m \lambda_e) \,. 
    \end{aligned}
\end{equation}
Under this ansatz, the free energy functional simplifies to
\begin{equation}
    F_{(q_e,q_m)}(\beta, N, \kappa) = N \left\{\frac{1}{\beta} \sum_{\alpha} \tr \log \left[-(\nabla - i\bar a_{\alpha})^2 + \frac{1}{4} + \mu_{\alpha}\right] - 2 \kappa (q_e \lambda_m + q_m \lambda_e)\right\} \,.
\end{equation}
This extremization problem can be solved for each choice of $q_e, q_m$. We now specialize to $(q_e, q_m) = (q,0)$ and $(q_e, q_m) = (q,q)$. The more general case contains no additional conceptual subtlety. 

\subsubsection{Leading order scaling dimension of \texorpdfstring{$\mathcal{M}_{(q,0)}$}{}}\label{app:Mq0_leading}

We first consider the flux configuration $(q_e, q_m) = (q,0)$. In the static background defined by \eqref{eq:saddle_general_ansatz}, the free energy functional simplifies to
\begin{equation}
    F_{(q,0)}(\beta, N, \kappa) = N \left\{\frac{1}{\beta} \sum_{\alpha} \tr \log \left[-(\nabla - i\bar a_{\alpha})^2 + \frac{1}{4} + \mu_{\alpha}\right] - 2 \kappa \lambda_m q\right\} \,.
\end{equation}
For fixed $\lambda_{\alpha}, \mu_{\alpha}$, the operator inside the trace-log can be diagonalized explicitly for each Matsubara frequency $\omega_n = 2\pi n/\beta$. Since the structure is different for $\alpha = e, m$, we state the results separately. In the $e$ sector, the scalar field sees a strength-$q$ monopole and the eigenfunctions are monopole harmonics:
\begin{equation}
    \left[-(\nabla - i \bar a_e)^2 + \frac{1}{4} + \mu_e\right] Y_{q,l,m}(\theta,\phi) e^{-i\omega_n \tau} = \left[(\omega_n - i \lambda_e)^2 + E_{l,q}(\mu_e)^2\right] Y_{q,l,m}(\theta,\phi) e^{-i\omega_n \tau} \,,
\end{equation}
where $l$ is the total angular momentum, $m$ is the azimuthal angular momentum, and $E_{l,q}(\mu)$ is the dispersion:
\begin{equation}
    E_{l,q}(\mu) = \sqrt{(l+1/2)^2 - q^2 + \mu} \,, \quad l = q,\, q+1, \ldots \,, \quad m = -l,\, - l + 1, \ldots, l \,. 
\end{equation}
On the other hand, for the $m$ sector, the scalar field sees no monopole and the eigenfunctions are regular spherical harmonics:
\begin{equation}
    \left[-(\nabla - i \bar a_m)^2 + \frac{1}{4} + \mu_m\right] Y_{0,l,m}(\theta,\phi) e^{-i\omega_n \tau} = \left[(\omega_n - i \lambda_m)^2 + E_{l,0}(\mu_m)^2\right] Y_{0,l,m}(\theta,\phi) e^{-i\omega_n \tau} \,. 
\end{equation}
Given these explicit spectra, the functional trace can be written as a sum over eigenvalues:
\begin{equation}
    \begin{aligned}
    \tr \log \left[-(\nabla - i\bar a_{e})^2 + \frac{1}{4} + \mu_{e}\right] &= \sum_{\omega_n} \sum_{l \geq q} d_l \log \left[(\omega_n - i \lambda_{e})^2 + E_{l,q}(\mu_{e})^2\right] \,, \\
    \tr \log \left[-(\nabla - i\bar a_{m})^2 + \frac{1}{4} + \mu_{m}\right] &= \sum_{\omega_n} \sum_{l \geq 0} d_l \log \left[(\omega_n - i \lambda_{m})^2 + E_{l,0}(\mu_{m})^2\right] \,, 
    \end{aligned}
\end{equation}
where $d_l = 2l+1$ is the degeneracy at fixed angular momentum $l$. For both $e$ and $m$ sectors, the sum over $\omega_n$ is UV divergent. To regularize it, we add and subtract a divergent piece as follows:
\begin{equation}
    \sum_{\omega_n} \log \left[(\omega_n - i \lambda)^2 + E^2\right] = \sum_{\omega_n} \log \left[\frac{(\omega_n - i \lambda)^2 + E^2}{(\omega_n - i \lambda)^2}\right] + \sum_{\omega_n} \log (\omega_n - i \lambda)^2
\end{equation}
The first sum is now convergent and can be evaluated explicitly. The second sum is divergent, but can be rewritten as
\begin{equation}
    \sum_{n \in \mathbb{Z}} \log (\omega_n - i\lambda)^2 = - \frac{d}{ds} \sum_{n \in \mathbb{Z}} (\omega_n - i\lambda)^{-2s} \bigg|_{s = 0} \,. 
\end{equation}
For sufficiently large $s$, the sum over $n$ is convergent and equal to the Hurwitz zeta function $\zeta(s, a)$ with $a =  \frac{-i \beta \lambda}{2\pi}$. Therefore, we can regularize the second term by analytically continuing $- \frac{d \zeta(s,a)}{ds}$ to $s = 0$. The final regularized result is given by
\begin{equation}
    \sum_{\omega_n} \log \left[(\omega_n - i \lambda)^2 + E^2\right] = \log \left(2 \cosh \beta E - 2 \cosh \beta \lambda \right) \,.
\end{equation}
Therefore, the full saddle point free energy functional can be written as a sum over the two sectors $F^{(0)}_{(q,0)}(\beta, \kappa) = F^{(0)}_{(q,0), e}(\beta, \kappa) + F^{(0)}_{(q,0), m}(\beta, \kappa)$ where
\begin{equation}
    \begin{aligned}
    F^{(0)}_{(q,0), e}(\beta, \kappa) &= \frac{1}{\beta} \sum_{l \geq q} d_l \log \left[2 \cosh \beta E_{l,q}(\mu_e) - 2 \cosh \beta \lambda_e \right] \,, \\
    F^{(0)}_{(q,0), m}(\beta, \kappa) &= \frac{1}{\beta} \sum_{l \geq 0} d_l \log \left[2 \cosh \beta E_{l,0}(\mu_m) - 2 \cosh \beta \lambda_m \right] - 2 \kappa \lambda_m q \,. 
    \end{aligned}
\end{equation}
In the $e$ sector, the saddle point free energy is identical to the saddle point free energy for a strength-$q$ monopole in scalar QED3 with no Chern-Simons term. As we know from Ref.~\cite{C210207377}, the stable saddle point solution is $\lambda_e = \mu_e = 0$ with a contribution to the scaling dimension $\Delta^{(0)}_{q, \rm SQED3}(\kappa = 0) \approx 0.12$. 

The $m$ sector is much more nontrivial. The first term is the Casimir energy of a free particle of mass $\mu_m$ on the sphere at a chemical potential $\lambda_m$. However, unlike the critical O(N) model where $\lambda_m = \mu_m = 0$, the presence of an additional $\kappa$-dependent term shifts the saddle point solutions for $\lambda_m, \mu_m$ and modifies the leading free energy, which we now calculate. The exact saddle point equations take the following form
\begin{equation}
    \begin{aligned}
    \frac{\delta F^{(0)}_{(q,0), m}}{\delta \lambda_m} &= \sum_{l \geq 0} \frac{-2 d_l \sinh \beta \lambda_m}{2 \cosh \beta E_{l,0}(\mu_m) - 2 \cosh \beta \lambda_m} - 2 \kappa q = 0 \,, \\
    \frac{\delta F^{(0)}_{(q,0), m}}{\delta \mu_m} &= \sum_{l \geq 0} \frac{\sinh \beta E_{l,0}(\mu_m)}{2 \cosh \beta E_{l,0}(\mu_m) - 2 \cosh \beta \lambda_m} \frac{d_l}{E_{l,0}(\mu_m)} = 0 \,.
    \end{aligned}
\end{equation}
In the large $\beta$ limit, up to errors that are exponentially suppressed in $\beta$, we can make the following approximation
\begin{equation}
    2\cosh \beta x \approx e^{\beta |x|} \,, \quad 2 \sinh \beta x \approx \sgn(x) e^{\beta |x|} \,. 
\end{equation}
Using this approximation, the saddle point equations simplify to
\begin{equation}
    \begin{aligned}
    2\kappa q \, \sgn(\lambda_m) &= \sum_{l \geq 0} \frac{d_l}{1 - e^{\beta E_{l,0}(\mu_m) - \beta |\lambda_m|}} \,, \quad \sum_{l \geq 0} \frac{1}{1 - e^{\beta |\lambda_m| - \beta E_{l,0}(\mu_m)}} \frac{d_l}{2E_{l,0}(\mu_m)} = 0 \,.
    \end{aligned}    
\end{equation}
For $\kappa > 0$, following related works in Ref.~\cite{C210207377}, the solution with minimal free energy corresponds to taking $\lambda_m \approx -E_{0,0}(\mu_m)$. This means we can set $\sgn(\lambda_m) = -1$ and obtain an explicit solution for $\lambda_m$ in terms of $\mu_m$ (valid up to errors that are exponentially suppressed in $\beta$):
\begin{equation}
    2 \kappa q = \frac{1}{e^{\beta \lambda_m + E_{0,0}(\mu_m)} - 1} \quad \rightarrow \quad \lambda_m = - E_{0,0}(\mu_m) - \frac{1}{\beta} \log \frac{2 \kappa q}{1 + 2 \kappa q} \,. 
\end{equation}
Plugging this solution into the saddle point equation for $\mu_m$, we get
\begin{equation}
    \frac{\kappa q}{E_{0,0}(\mu_m)} + \sum_{l \geq 1} \frac{d_l}{2 E_{l,0}(\mu_m)} = 0 \,. 
\end{equation}
The sum over $l$ is again UV divergent and can be regularized using the Hurwitz zeta function. After regularization, we obtain a convergent equation
\begin{equation}
    \frac{\kappa q}{E_{0,0}(\mu_m)} + \sum_{l \geq 1} \left(\frac{d_l}{2 E_{l,0}(\mu_m)} - 1 \right) = 0 \,. 
\end{equation}
We were not able to obtain a closed form solution for $\mu_m$ as a function of $\kappa$. However, for each $\kappa$, one can solve the equation numerically, and plug it back into the free energy to obtain:
\begin{equation}
    \begin{aligned}
    F^{(0)}_{(q,0), m}(\beta, \kappa) &\approx \frac{1}{\beta} \sum_{l \geq 0} d_l \log \left[e^{\beta E_{l,0}(\mu_m)}\right] + \frac{1}{\beta} \sum_{l \geq 0} d_l \log \left[1 - e^{-\beta \lambda_m - \beta E_{l,0}(\mu_m)}\right] + 2 \kappa q \left[E_{0,0}(\mu_m) + \frac{1}{\beta} \log \frac{2 \kappa q}{1 + 2 \kappa q}\right] \\
    &= \sum_{l \geq 0} d_l E_{l,0}(\mu_m) + 2 \kappa q E_{0,0}(\mu_m) + \mathcal{O}(\beta^{-1}) \,.
    \end{aligned}
\end{equation}
The sum over $l$ can again be regularized by evaluating the divergent piece via zeta-function regularization:
\begin{equation}
    2 \lim_{s \rightarrow 0} \sum_{l\geq 0} (l+1/2) \left[(l+1/2)^{1-2s} + (1/2-s) (l+1/2)^{-1-2s} \mu_m \right] = 2 \zeta(-2, 1/2) + \mu_m  \zeta(0,1/2) = 0 \,.
\end{equation}
The remaining convergent sum can be evaluated numerically. After taking a $\beta \rightarrow \infty$ limit, we recover the leading contribution to the monopole scaling dimension
\begin{equation}\label{eq:Delta0_q0_final}
    \Delta^{(0)}_{(q,0)}(\kappa) = \Delta^{(0)}_{q, \rm SQED3}(\kappa = 0) + \Delta^{(0)}_{q, m}(\kappa)  \,, 
\end{equation}
where
\begin{equation}
    \Delta^{(0)}_{q, m}(\kappa) = 2 \kappa q E_{0,0}(\mu_m) + 2 \sum_{l\geq 0} (l+1/2) \left\{\left[(l+1/2)^2 + \mu_m\right]^{1/2} - (l+1/2) - \frac{1}{2} (l+1/2)^{-1} \mu_m \right\} \,.
\end{equation}
This equation can then be used to generate the plot in Fig.~\ref{fig:monopole}. 

Further analytic progress can be made in the large $\kappa$ limit. In this limit, the saddle point equation can be approximated as an integral
\begin{equation}
    \frac{\kappa q}{\sqrt{\mu_m}} \approx \int_1^{\infty} dl \left(1 - \frac{l+1/2}{\sqrt{(l+1/2)^{2} + \mu_m}} \right) = \frac{1}{2} (\sqrt{9 + 4 \mu_m} - 3) \,.
\end{equation}
Therefore, in the large $\kappa$ limit, the leading order solution for $\mu_m, \lambda_m$ is given by
\begin{equation}
    \mu_m(\kappa) = \kappa q \,, \quad \lambda_m \approx - E_{0,0}(\mu_m) \approx - \sqrt{\kappa q} \,. 
\end{equation}
Using this solution, we can evaluate $F^{(0)}_{(q,0), m}(\beta, \kappa)$ using $\zeta$-regularization:
\begin{equation}
    \begin{aligned}
    F^{(0)}_{(q,0), m}(\beta, \kappa) &\approx 2 (\kappa q)^{3/2} + 2 \int_0^{\infty} dl (l+1/2) \left\{\left[(l+1/2)^2 + \kappa q\right]^{1/2}  - (l+1/2) - \frac{\kappa q}{2 (l+1/2)}\right\} \\
    &= 2 (\kappa q)^{3/2} + \frac{1 + 6 \kappa q - (1+ 4\kappa q )^{3/2}}{12} \approx \frac{4}{3} (\kappa q)^{3/2} \,. 
    \end{aligned}
\end{equation}

\subsubsection{Leading order scaling dimension of \texorpdfstring{$\mathcal{M}_{(q,q)}$}{}}

Now let us consider a duality-symmetric flux configuration $(q_e, q_m) = (q,q)$. Assuming that the duality symmetry is unbroken at the critical point, the saddle point ansatz also needs to respect the duality symmetry and takes the form $\bar a_{e,0} = \bar a_{m,0} = \bar a, \bar \mu_e = \bar \mu_m = \mu$, where 
\begin{equation}
    \bar a_0(\theta, \phi) = -i \lambda \,, \quad d \bar a(\theta, \phi) = q \sin \theta d \theta \wedge d \phi \,.
\end{equation}
With this ansatz, the saddle point action simplifies tremendously and becomes identical to the saddle point action for scalar QED3 with $2N$ flavors of scalar fields and a self Chern-Simons term at level $2k = 2\kappa N$:
\begin{equation}
    S_{\rm saddle} = 2N \left\{ \tr \log \left[(i\nabla - \bar a)^2 + \frac{1}{4} + \mu\right] - 2 \kappa \lambda q \right] \,. 
\end{equation}
Therefore, the strength $(q,q)$ monopole in the self-dual theory with fixed $\kappa$ has the same leading order scaling dimension as the strength-q monopole in the theory of $2N$ complex scalars coupled to a self Chern-Simons term at level $2k$. This relation is summarized by a simple equation: 
\begin{equation}
    \Delta^{(0)}_{(q,q)}(\kappa) = 2 \Delta^{(0)}_{q, \rm SQED3}(\kappa) \,.
\end{equation}

\subsection{Subleading order: quadratic fluctuations around the saddle point}\label{app:monopole_subleading}

Having finished the saddle point analysis, we move on to the subleading corrections given by quadratic fluctuations of the gauge and matter fields around the saddle point. To that end, we expand the original fields as $a_{\alpha} \rightarrow \bar a_{\alpha} + a_{\alpha}, \mu_{\alpha} \rightarrow \bar \mu_{\alpha} + i \sigma_{\alpha}$ where $\bar a_{\alpha}, \bar \mu_{\alpha}$ are the saddle point solutions. After integrating over the scalar field, the effective action for $a_{\alpha}, \sigma_{\alpha}$ takes the form
\begin{equation}
    \begin{aligned}
    S &= N \sum_{\alpha = e, m} \tr \log \left[-(\nabla - i \bar a_{\alpha} - i a_{\alpha})^2 + \frac{1}{4} + \bar \mu_{\alpha} + i \sigma_{\alpha} \right] - \frac{ik}{2\pi} \int d^3 x (\bar a_e + a_e) d (\bar a_m + a_m) \,.
    \end{aligned}
\end{equation}
Now let us define the matter field kernel $G_{\alpha}^{-1} = - (\nabla - i \bar a_{\alpha})^2 + \frac{1}{4} + \bar \mu_{\alpha}$. Expanding the trace log terms to quadratic order, we find that the matter field contribution to the quadratic action takes the form
\begin{equation}
    S_{2, \phi} = \sum_{\alpha = e, m} S_{2, \alpha} \,,
\end{equation}
where
\begin{equation}
    \begin{aligned}
        S_{2, \alpha} &= N \tr \log \left[G_{\alpha}^{-1} + i D^{\mu} a_{\alpha, \mu} + i a_{\alpha,\mu} D^{\mu} + a_{\alpha}^2 + i \sigma_{\alpha}\right] - N \tr \log G_{\alpha}^{-1} \\
        &= N \tr \log \left[1 + i G_{\alpha} D^{\mu} a_{\alpha, \mu} + i G_{\alpha} a_{\alpha,\mu} D^{\mu} + G_{\alpha} a_{\alpha}^2 + i G_{\alpha} \sigma_{\alpha}\right] \\
        &\approx N \tr G_{\alpha} a_{\alpha}^2 + \frac{N}{2} \tr (G_{\alpha} D^{\mu} a_{\alpha, \mu} + G_{\alpha} a_{\alpha,\mu} D^{\mu} + G_{\alpha} \sigma_{\alpha}) (G_{\alpha} D^{\nu} a_{\alpha, \nu} + G_{\alpha} a_{\alpha,\nu} D^{\nu} + G_{\alpha} \sigma_{\alpha}) \,.
    \end{aligned}
\end{equation}
After some simple algebra, one can write the quadratic action more compactly as
\begin{equation}
    S_{2,\alpha} = \frac{N}{2} \sum_{\alpha} \int \left[a_{\alpha, \mu}(x), \sigma_{\alpha}(x)\right] \begin{pmatrix}
        K_{\alpha}^{\mu\nu}(x,y) & K_{\alpha}^{\mu\sigma}(x,y) \\ K_{\alpha}^{\sigma \nu}(x,y)  & K_{\alpha}^{\sigma\sigma}(x,y) 
    \end{pmatrix} \begin{pmatrix}
        a_{\alpha, \nu}(y) \\ \sigma_{\alpha}(y)
    \end{pmatrix}  \,,
\end{equation}
where the kernels are given by 
\begin{equation}
    \begin{aligned}
    K^{\mu\nu}_{\alpha}(x,y) &= \left[D^{\mu}_{\alpha, x} G_{\alpha}(y,x)\right] \left[D^{\nu}_{\alpha,y} G_{\alpha}(x,y)\right] - G_{\alpha}(x,y) \left[D^{\mu}_{\alpha,x} D^{\nu}_{\alpha,y} G_{\alpha}(y,x)\right] \\
    &\hspace{2cm}+ \left[D^{\mu}_{\alpha,x} G_{\alpha}(x,y)\right] \left[D^{\nu}_{\alpha,y} G_{\alpha}(y,x)\right] - G_{\alpha}(y,x) \left[D^{\mu}_{\alpha,x} D^{\nu}_{\alpha,y} G_{\alpha}(x,y)\right] + 2 G_{\alpha}(x,y) g^{\mu\nu} \delta(x-y)  \,, \\
    K^{\sigma \nu}_{\alpha}(x,y) &= G_{\alpha}(x,y) D^{\nu}_{\alpha,y} G_{\alpha}(y,x) - G_{\alpha}(y,x) D^{\nu}_{\alpha,y} G_{\alpha}(x,y) \,, \\
    K^{\sigma\sigma}_{\alpha}(x,y) &= G_{\alpha}(x,y) G_{\alpha}(y,x) \,,
    \end{aligned}
\end{equation}
with $D^{\mu}_{\alpha,x}$ denoting the covariant derivative $\partial^{\mu}_x - i \bar a_{\alpha}^{\mu}(x)$ defined at the $\alpha$-sector saddle point. Combining the matter field quadratic action with the mutual Chern-Simons term 
\begin{equation}
    S_{2, CS} = - \frac{ik}{2\pi} \int d^3 x a_e d a_m \,, 
\end{equation}
we can write the full quadratic action as
\begin{equation}
    S = \frac{N}{2} \int [a_e, a_m, \sigma_e, \sigma_m] K [a_e, a_m, \sigma_e, \sigma_m]^T \,, 
\end{equation}
where the extended kernel can be expressed in terms of the previously defined subkernels:
\begin{equation}
    K_{(q_e,q_m)} = \begin{pmatrix}
        K^{\mu\nu}_e & K_{CS} & K^{\mu \sigma}_e & 0 \\ K_{CS} & K^{\mu\nu}_m & 0 & K^{\mu \sigma}_m \\
        K^{\sigma \nu}_e & 0 & K^{\sigma\sigma}_e & 0 \\ 
        0 & K^{\sigma \nu}_m & 0 & K^{\sigma\sigma}_m 
    \end{pmatrix} \,.
\end{equation}
Due to the block structure of $K_{(q_e,q_m)}$, we can further simplify its determinant to
\begin{equation}\label{eq:Kqeqm_generalform}
    \begin{aligned}
    \det K_{(q_e,q_m)} &= \det \begin{pmatrix}
        K^{\sigma\sigma}_e & 0 \\ 
        0 & K^{\sigma\sigma}_m 
    \end{pmatrix} \cdot \det \left[\begin{pmatrix}
        K^{\mu\nu}_e & K_{CS} \\ K_{CS} & K^{\mu\nu}_m 
    \end{pmatrix} - \begin{pmatrix}
        K^{\mu \sigma}_e & 0 \\ 0 & K^{\mu \sigma}_m
    \end{pmatrix} \begin{pmatrix}
        (K^{\sigma\sigma}_e)^{-1} & 0 \\ 0 & (K^{\sigma\sigma}_m)^{-1} 
    \end{pmatrix} \begin{pmatrix}
        K^{\sigma \nu}_e & 0 \\ 0 & K^{\sigma \nu}_m 
    \end{pmatrix} \right] \\
    &= \det K^{\sigma \sigma}_e \cdot \det K^{\sigma\sigma}_m \cdot \det \begin{pmatrix}
        K^{\mu\nu}_e - K^{\mu \sigma}_e (K^{\sigma\sigma}_e)^{-1} K^{\sigma \nu}_e & K_{CS} \\ K_{CS} & K^{\mu\nu}_m - K^{\mu \sigma}_m (K^{\sigma\sigma}_m)^{-1} K^{\sigma \nu}_m 
    \end{pmatrix} \,.
    \end{aligned}
\end{equation}
From here, one can obtain a formal expression for the subleading free energy 
\begin{equation}
    F^{(1)}_{(q_e,q_m)}(\beta, \kappa) = \frac{1}{\beta} \log \det \frac{K_{(q_e,q_m)}(\beta, \kappa)}{K_{(0,0)}(\beta, \kappa)} \quad \rightarrow \quad \Delta^{(1)}_{(q_e, q_m)}(\kappa) = \lim_{\beta \rightarrow \infty} F^{(1)}_{(q_e,q_m)}(\beta, \kappa) \,,
\end{equation}
where a factor of $K_{(0,0)}(\beta, \kappa)$ is added to regularize the determinant. In the following two sections, we make some general analytic remarks about the functional determinant. For the $\mathcal{M}_{(q,0)}$ monopole, we will argue that the expression for $\Delta^{(1)}_{(q,0)}$ simplifies dramatically in the large $\kappa$ limit and scales as $\kappa^{3/2}$. For the $\mathcal{M}_{(q,q)}$ monopole, we will prove that the quadratic fluctuation correction $\Delta^{(1)}_{(q,q)}(\kappa)$ is twice as large as the quadratic fluctuation correction $\Delta^{(1)}_{q, \rm SQED3}(\kappa)$ for a strength-$q$ monopole in scalar QED3 with self-Chern-Simons level $k = N \kappa$. The precise numerical evaluation of $\Delta^{(1)}_{(q_e,q_m)}(\kappa)$ for general $q_e, q_m, \kappa$ is orthogonal to the conceptual theme of this work and will be deferred to the future. 

\subsubsection{Subleading scaling dimension of \texorpdfstring{$\mathcal{M}_{(q,0)}$}{}}\label{app:Mq0_subleading}

The goal of this section is to understand the structure of $\Delta^{(1)}_{(q,0)}(\kappa)$ at large $\kappa$ and show that it scales as $\kappa^{3/2}$ in the $\kappa \rightarrow \infty$ limit. For that purposes, we first recall the saddle point solution:
\begin{equation}
    \lambda_e = \mu_e = 0 \,, \quad \lambda_m \approx - \sqrt{\mu_m} \sim \sqrt{\kappa} \,. 
\end{equation}
From this solution, we immediately see that all components of the kernel that involve the $e$ sector do not depend on $\kappa$. On the other hand, the Chern-Simons kernel $K_{\rm CS}$ is proportional to $\kappa$. 
\begin{comment}
\begin{equation}
    K^{\mu\nu}_e = K^{\mu\sigma}_e = K^{\sigma\sigma}_e = \mathcal{O}(1) \,, \quad K^{\mu\nu}_m = \mathcal{O}(\kappa^{-1}) \,, \quad K^{\mu\sigma}_m = \mathcal{O}(\kappa^{-3/2}) \,, \quad K^{\sigma\sigma}_m = \mathcal{O}(\kappa^{-2}) \,, \quad K_{CS} = \mathcal{O}(\kappa) \,. 
\end{equation}
This scaling immediately implies that in the large $\kappa$ limit, 
\begin{equation}
    K^{\mu\nu}_e - K^{\mu \sigma}_e (K^{\sigma\sigma}_e)^{-1} K^{\sigma \nu}_e = \mathcal{O}(1) \,, \quad K^{\mu\nu}_m - K^{\mu \sigma}_m (K^{\sigma\sigma}_m)^{-1} K^{\sigma \nu}_m = \mathcal{O}(\kappa^{-1}) \,. 
\end{equation}
\end{comment}
As a result, the off-diagonal $K_{CS}$ blocks in \eqref{eq:Kqeqm_generalform} dominate over the diagonal blocks and the functional determinant simplifies to
\begin{equation}
    \det K_{(q,0)} \approx \det K^{\sigma\sigma}_{e, q} \cdot \det K^{\sigma\sigma}_{m, q} \cdot \det K_{CS}^2 
\end{equation}
which immediately implies that
\begin{equation}
    \Delta^{(1)}_{(q,0)}(\kappa) \approx \lim_{\beta \rightarrow \infty} \frac{1}{\beta} \left[\log \frac{ \det K_{(q,0)}}{\det K_{(0,0)}}\right] = \lim_{\beta \rightarrow \infty} \frac{1}{\beta} \left[\log \frac{\det K^{\sigma\sigma}_{e, q} \cdot \det K^{\sigma\sigma}_{m, q} \cdot \det K_{CS}^2}{\det K^{\sigma\sigma}_{e, q=0} \cdot \det K^{\sigma\sigma}_{m, q=0} \cdot \det K_{CS}^2}\right] \,. 
\end{equation}
Since $K_{\rm CS}$ is independent of $q$ and the scalar kernel for $q = 0$ is independent of the flavor index $\alpha = e,m$, the above expression further reduces to
\begin{equation}
    \Delta^{(1)}_{(q,0)}(\kappa) \approx \lim_{\beta \rightarrow \infty} \frac{1}{\beta} \left[\tr \log \left(\frac{K^{\sigma\sigma}_{e,q}}{K^{\sigma\sigma}_{0}}\right) + \tr \log \left(\frac{K^{\sigma\sigma}_{m,q}}{K^{\sigma\sigma}_{0}}\right) \right] \,,
\end{equation}
where $K^{\sigma\sigma}_{0} \equiv K^{\sigma\sigma}_{e, q=0} = K^{\sigma\sigma}_{m, q=0}$.
Let us now estimate these two terms in turn. The first step is to derive some explicit formulae for the saddle point Green's function of each matter field. Using the spectra computed before, we can expand the matter propagators in monopole harmonics:
\begin{equation}
    \begin{aligned}
    G_e(\tau, \tau', \theta, \theta', \phi, \phi') &= \frac{1}{\beta} \sum_n \sum_{l \geq q, m}  \frac{e^{-i \omega_n (\tau - \tau')}}{\omega_n^2 + E_{l,q}(0)^2} Y_{q,l,m}(\theta,\phi) Y_{q,l,m}^*(\theta', \phi') \,, \\
    G_m(\tau, \tau', \theta, \theta', \phi, \phi') &= \frac{1}{\beta} \sum_n \sum_{l \geq 0, m}  \frac{e^{-i \omega_n (\tau - \tau')}}{(\omega_n - i \lambda_m)^2 + E_{l,0}(\mu_m)^2} Y_{0,l,m}(\theta,\phi) Y_{0,l,m}^*(\theta', \phi') \,.
    \end{aligned}
\end{equation}
The Matsubara sums can be evaluated exactly using the Poisson resummation formula
\begin{equation}
    \frac{1}{\beta} \sum_n  \frac{e^{-i \omega_n (\tau - \tau')}}{(\omega_n - i a)^2 + b^2} = \frac{e^{a(\tau-\tau')}}{2b} \left[e^{-b |\tau-\tau'|} + \frac{e^{-b(\tau-\tau')}}{e^{\beta (b-a)} - 1} + \frac{e^{b(\tau-\tau')}}{e^{\beta (b+a)} - 1} \right] \,.
\end{equation}
After plugging in the saddle point values for $\lambda_e, \mu_e$, expressing $\lambda_m$ in terms of $\mu_m$ (which is a numerically determined function of $\kappa$), and dropping terms that are exponentially suppressed in $\beta$, the matter field propagators reduce to
\begin{equation}
    \begin{aligned}
    G_e(\tau, \tau', \theta, \theta', \phi, \phi') &= \sum_{l \geq q, m} \frac{1}{2 E_{l,q}(0)} \left[e^{-E_{l,q}(0) |\tau - \tau'|} + e^{-E_{l,q}(0) (\tau - \tau' + \beta)} + e^{-E_{l,q}(0) (-\tau + \tau' + \beta)} \right] Y_{q,l,m}(\theta,\phi) Y_{q,l,m}^*(\theta', \phi') \,, \\
    G_m(\tau, \tau', \theta, \theta', \phi, \phi') &= \frac{e^{\lambda_m(\tau-\tau')}}{2 E_{0,0}(\mu_m)} \left[e^{-E_{0,0}(\mu_m) |\tau - \tau'|} + e^{- E_{0,0}(\mu_m) (\tau-\tau'+\beta)} + \frac{e^{E_{0,0}(\mu_m) (\tau - \tau')}}{e^{\beta(E_{0,0}(\mu_m) + \lambda_m)} - 1}\right]Y_{0,0,m}(\theta,\phi) Y_{0,0,m}^*(\theta', \phi') \\
    &\hspace{2cm} + \sum_{l \geq 1, m} \frac{e^{\lambda_m(\tau-\tau')}}{2 E_{l,0}(\mu_m)} e^{-E_{l,0}(\mu_m) |\tau - \tau'|} Y_{0,l,m}(\theta,\phi) Y_{0,l,m}^*(\theta', \phi') \,.
    \end{aligned}
\end{equation}

\begin{center}
    \textbf{Calculations in the e sector} 
\end{center}

\noindent First, we consider the e sector, where $\lambda_e = \mu_e = 0$ but $q \neq 0$. Here, we will need the monopole harmonics summation identity
\begin{equation}
    \sum_{m=-l}^l Y_{qlm}(\theta,\phi) Y^*_{qlm}(\theta',\phi') = e^{-2iq \Theta} F_{q,l}(\gamma) \,,
\end{equation}
where 
\begin{equation}
    F_{q,l}(\gamma) = \sqrt{\frac{2l+1}{4\pi}} Y_{q,l,-q}(\gamma,0) \,, \quad e^{i\Theta} \cos (\gamma/2) = \cos(\theta/2) \cos(\theta'/2) + e^{-i(\phi-\phi')} \sin (\theta/2) \sin (\theta'/2)
\end{equation}
Using these identities and taking the $\beta = \infty$ limit, we find 
\begin{equation}
    \begin{aligned}
    G_e(x,x') &= \sum_{l \geq q, m} \frac{Y_{qlm}(\theta,\phi) Y^*_{qlm}(\theta',\phi')}{2 E_{l,q}(0)} e^{-E_{l,q}(0) |\tau-\tau'|} = \sum_{l \geq q} e^{-2iq\Theta} F_{q,l}(\gamma) \frac{e^{-E_{l,q}(0) |\tau-\tau'|}}{2 E_{l,q}(0)}
    \end{aligned}
\end{equation}
From here, we can compute the Fourier representation of the kernel
\begin{equation}
    \begin{aligned}
        K^{\sigma\sigma}_{e,l}(\omega_n) &= \frac{4\pi}{2l+1} \int d\tau \sin \theta d \theta d \phi e^{i\omega_n \tau} \sum_{m=-l}^l Y^*_{lm}(x) K^{\sigma\sigma}_{e}(x,x') Y_{lm}(x') \bigg|_{x'=0}\\
        &= \frac{4\pi \cdot 2\pi}{2l+1} \sum_{l_1,l_2 \geq q} \int d\tau e^{i\omega_n \tau} \frac{e^{-E_{l_1,q}(0) |\tau| - E_{l_2,q}(0) |\tau|}}{4 E_{l_1,q}(0) E_{l_2,q,0}(0)} \left[\int d \theta \sin \theta F_{0,l}(\theta) F_{q,l_1}(\theta) F_{q,l_2}(\theta)\right] \\
        &= \frac{8\pi^2}{2l+1} \sum_{l_1, l_2 \geq q} \frac{E_{l_1,q}(0) + E_{l_2,q}(0)}{2 E_{l_1,q}(0) E_{l_2,q}(0) \left\{\omega^2 + \left[E_{l_1,q}(0) + E_{l_2,q}(0)\right]^2\right\}} \int d \theta \sin \theta F_{0,l}(\theta)F_{q,l_1}(\theta) F_{q,l_2}(\theta) \,. 
    \end{aligned}
\end{equation}
Now let us recall the more general triple product identity for Jacobi polynomials:
\begin{equation}\label{eq:triple_prod_identity}
    \int d \theta \sin \theta F_{0,l}(\theta)F_{q,l_1}(\theta) F_{q,l_2}(\theta) = \frac{(2l+1)(2l_1+1)(2l_2+1)}{32 \pi^3} \begin{pmatrix}
        l & l_1 & l_2 \\ 0 & -q & q 
    \end{pmatrix}^2 \,. 
\end{equation}
Using this triple product identity, $K^{\sigma\sigma}_{e,l}(\omega_n)$ can be represented as a convergent sum:
\begin{equation}
    K^{\sigma\sigma}_{e,l}(\omega_n) = \sum_{l_1, l_2 \geq q} \frac{(2l_1+1)(2l_2+1)}{8\pi E_{l_1,q}(0) E_{l_2,q}(0)} \frac{E_{l_1,q}(0) + E_{l_2,q}(0)}{\omega^2 + \left[E_{l_1,q}(0) + E_{l_2,q}(0)\right]^2} \begin{pmatrix}
        l & l_1 & l_2 \\ 0 & -q & q 
    \end{pmatrix}^2  \,.
\end{equation}
When $q = 0$, we recover the analytic answer:
\begin{equation}
    K^{\sigma\sigma}_{e,l}(\omega_n) \rightarrow \frac{1}{2\pi} \sum_{l_1, l_2 \geq 0} \frac{l_1+l_2+1}{\omega^2 + (l_1+l_2+1)^2} \begin{pmatrix}
        l & l_1 & l_2 \\ 0 & 0 & 0
    \end{pmatrix}^2 = \left|\frac{\Gamma\left(\frac{l+1+i\omega}{2}\right)}{4 \Gamma\left(\frac{l+2+i\omega}{2}\right)}\right|^2 \,.
\end{equation}
For general $q$, the sum needs to be done numerically. However, note that the answer does not depend on $\kappa$. Therefore, the contribution will be an $\mathcal{O}(1)$ term that competes with terms that we have already dropped. 

\begin{center}
    \textbf{Calculations in the m sector} 
\end{center}

\noindent Finally, we treat the $m$ sector, where $q = 0$ and $\lambda_m, \mu_m \neq 0$:
\begin{equation}
    \begin{aligned}
    G_m(x,x') &= e^{\lambda(\tau-\tau')}\sum_{l,m} \frac{Y_{lm}(\theta,\phi) Y^*_{lm}(\theta',\phi')}{2 E_{l,0}(\mu)} \left(e^{- E_{l,0}(\mu)|\tau-\tau'|} + \frac{e^{-E_{l,0}(\mu)(\tau-\tau')}}{e^{\beta E_{l,0}(\mu) - \beta \lambda} - 1} + \frac{e^{E_{l,0}(\mu)(\tau-\tau')}}{e^{\beta E_{l,0}(\mu) + \beta \lambda} - 1}\right) \\
    &= e^{\lambda(\tau-\tau')} \sum_l \frac{2l+1}{8\pi \cdot E_{l,0}(\mu)} P_l(\cos \gamma) \left(e^{- E_{l,0}(\mu)|\tau-\tau'|} + \frac{e^{-E_{l,0}(\mu)(\tau-\tau')}}{e^{\beta E_{l,0}(\mu) - \beta \lambda} - 1} + \frac{e^{E_{l,0}(\mu)(\tau-\tau')}}{e^{\beta E_{l,0}(\mu) + \beta \lambda} - 1}\right) \,.
    \end{aligned}
\end{equation}
Plugging in the saddle point solution
\begin{equation}
    \lambda = - E_{l,0}(\mu) + \beta^{-1} \log \frac{1 + 2 \kappa q}{2 \kappa q} \,,
\end{equation}
we can obtain the m-sector scalar Green's function to exponential accuracy in $\beta$:
\begin{equation}
    \begin{aligned}
    G_m(x,x') &\approx e^{\lambda(\tau-\tau')} \left[\sum_l \frac{2l+1}{8\pi \cdot E_{l,0}(\mu)} P_l(\cos \gamma) e^{- E_{l,0}(\mu)|\tau-\tau'|} + \frac{1}{8\pi \cdot E_{0,0}(\mu)} \frac{e^{E_{0,0}(\mu) (\tau - \tau')}}{\frac{1 + 2 \kappa q}{2\kappa q} - 1}\right] \\
    &= e^{\lambda(\tau-\tau')} \left[G_{m,\rm CP}(x, x') + e^{E_{0,0}(\mu) (\tau - \tau')} G_{m, \rm NCP}(x, x') \right] \,,
    \end{aligned}
\end{equation}
where we defined the CP-invariant/non-CP-invariant kernels $G_{m,\rm CP}$/$G_{m,\rm NCP}$:
\begin{equation}
    G_{m,\rm CP}(x,x') = \sum_l \frac{2l+1}{8\pi E_{l,0}(\mu)} P_l(\cos \gamma) e^{- E_{l,0}(\mu)|\tau-\tau'|} \,, \quad G_{m,\rm NCP}(x,x') = \frac{2 \kappa q}{8\pi E_{0,0}(\mu)} \,. 
\end{equation}
Note that both kernels are invariant under $x \leftrightarrow x'$. However, the NCP kernel is multiplied by an extra phase factor which is not invariant under CP. From here, it follows that
\begin{equation}
    \begin{aligned}
    K^{\sigma\sigma}_m(x,x') &= G_m(x,x') G_m(x',x) \\
    &= G_{m,\rm CP}(x, x')^2 + G_{m, \rm NCP}(x,x')^2 + G_{m, \rm NCP}(x,x') G_{m, \rm CP}(x',x) \left[e^{E_{0,0}(\mu) (\tau - \tau')} + e^{-E_{0,0}(\mu) (\tau - \tau')}\right]  \,.
    \end{aligned}
\end{equation}
The term that that contains only NCP kernels is just a constant. Upon taking the Fourier transform, we get a linear divergence in $\beta$ which does not contribute to the scaling dimension. The other two terms are more interesting. Let us first evaluate the CP-invariant term
\begin{equation}
    \begin{aligned}
    K^{(\rm CP)}_{m,l}(\omega) &= \frac{4\pi}{2l+1} \int d^3 x \sqrt{g} e^{i\omega(\tau-\tau')} \sum_{m=-l}^l Y^*_{lm}(x) K^{\sigma\sigma}_{m, \rm CP}(x,x') Y_{lm}(x') \bigg|_{x'=0} \\
    &= \frac{4\pi \cdot 2\pi}{2l+1} \int d \tau \sin \theta d \theta e^{i\omega \tau} K^{\sigma\sigma}_{m, \rm CP}(\tau, \theta, \phi,0) F_{0,l}(\theta) \\
    &= \sum_{l_1, l_2} \frac{8\pi^2}{4(2l+1) E_{l_1,0}(\mu) E_{l_2,0}(\mu)} \int d \tau e^{i\omega \tau} e^{i\omega \tau - E_{l_1,0}(\mu) |\tau| - E_{l_2,0}(\mu) |\tau|} \left[\int d \theta \sin \theta F_{0,l}(\theta) F_{0,l_1}(\theta) F_{0,l_2}(\theta)\right] \,.
    \end{aligned}
\end{equation}
The integral over $\tau$ is straightforward:
\begin{equation}
    \int_{-\infty}^{\infty} d \tau e^{i\omega \tau} e^{- E_{l_1,0}(\mu)|\tau| - E_{l_2,0}(\mu)|\tau|} = 2 \frac{E_{l_1,0}(\mu) + E_{l_2,0}(\mu)}{\omega^2 + \left[E_{l_1,0}(\mu) + E_{l_2,0}(\mu)\right]^2} \,.
\end{equation}
The integral over $\theta$ can again be done using the triple product identity in \eqref{eq:triple_prod_identity} and the final answer is
\begin{equation}
    \begin{aligned}
        K^{(\rm CP)}_{m,l}(\omega) &= \sum_{l_1,l_2} \frac{4\pi^2 (2l_1+1)(2l_2+1)}{32 \pi^3 E_{l_1,0}(\mu) E_{l_2,0}(\mu)} \frac{E_{l_1,0}(\mu) + E_{l_2,0}(\mu)}{\omega^2 + \left[E_{l_1,0}(\mu) + E_{l_2,0}(\mu)\right]^2} \begin{pmatrix}
            l & l_1 & l_2 \\ 0 & 0 & 0
        \end{pmatrix}^2 \\
        &= \sum_{l_1, l_2} \frac{(2l_1+1)(2l_2+1)}{8\pi E_{l_1,0}(\mu) E_{l_2,0}(\mu)} \frac{E_{l_1,0}(\mu) + E_{l_2,0}(\mu)}{\omega^2 + \left[E_{l_1,0}(\mu) + E_{l_2,0}(\mu)\right]^2} \begin{pmatrix}
        l & l_1 & l_2 \\ 0 & 0 & 0
    \end{pmatrix}^2  \,.
    \end{aligned}
\end{equation}
The situation is simpler for the non-CP preserving kernel:
\begin{equation}
    \begin{aligned}
    K^{(\rm NCP)}_{m,l}(\omega) &= \int d \tau \sin \theta d \theta d \phi e^{i\omega \tau} 2 \cosh \left[E_{0,0}(\mu) \tau\right] G_{m, \rm NCP}(\tau, 0) G_{m, \rm CP}(\tau,0) P_l(\cos \theta) \\
    &= \frac{\kappa q}{2 E_{0,0}(\mu)} \int d \tau \sin \theta d \theta e^{i\omega \tau} 2 \cosh \left[E_{0,0}(\mu) \tau\right] P_l(\cos \theta) \sum_{l_1} \frac{2l_1+1}{8\pi E_{l_1,0}(\mu)} P_{l_1}(\cos \theta) e^{-E_{l_1,0}(\mu) |\tau|} 
    \end{aligned}
\end{equation}
Now let us recall the simpler angular integral
\begin{equation}
    \int \sin \theta d \theta P_{l_1}(\cos \theta) P_{l_2}(\cos \theta) = \frac{2}{2l_1+1} \delta_{l_1 l_2} \,, 
\end{equation}
and the temporal integration
\begin{equation}
    \int_0^{\infty} d \tau e^{i\omega \tau} \cdot 2 \cosh \left[E_{0,0}(\mu) \tau \right] e^{-E_{l,0}(\mu) |\tau|} = \begin{cases}
        2 \pi \delta(\omega) & l = 0 \\ 2 \left(\frac{E_{0,0}(\mu) + E_{l,0}(\mu)}{\omega^2+\left[E_{0,0}(\mu) + E_{l,0}(\mu)\right]^2} + \frac{E_{l,0}(\mu) - E_{0,0}(\mu)}{\omega^2+\left[E_{l,0}(\mu) - E_{0,0}(\mu)\right]^2}\right) & l \geq 1 
    \end{cases} \,. 
\end{equation}
Plugging these formulae into $K^{(\rm NCP)}_{m,l}(\omega)$ and throwing away the $\delta$-function contributions, we find that 
\begin{equation}
    \begin{aligned}
        K^{(\rm NCP)}_{m,l}(\omega) &= \frac{\kappa q (2l+1)}{16 \pi E_{0,0}(\mu) E_{l,0}(\mu)} \cdot \frac{2}{2l+1} \cdot 2 \left\{\frac{E_{0,0}(\mu) + E_{l,0}(\mu)}{\omega^2+\left[E_{0,0}(\mu) + E_{l,0}(\mu)\right]^2} + \frac{E_{l,0}(\mu) - E_{0,0}(\mu)}{\omega^2+\left[E_{l,0}(\mu) - E_{0,0}(\mu)\right]^2}\right\} \\
        &= \frac{\kappa q}{4 \pi E_{0,0}(\mu) E_{l,0}(\mu)} \left\{\frac{E_{0,0}(\mu) + E_{l,0}(\mu)}{\omega^2+\left[E_{0,0}(\mu) + E_{l,0}(\mu)\right]^2} + \frac{E_{l,0}(\mu) - E_{0,0}(\mu)}{\omega^2+\left[E_{l,0}(\mu) - E_{0,0}(\mu)\right]^2}\right\} \,.
    \end{aligned}
\end{equation}
\begin{comment}
The large $l$ and large $\omega$ asymptotics of this kernel can be easily worked out:
\begin{equation}
    K^{(\rm NCP)}_{m,l}(\omega) \approx_{\beta, l \rightarrow \infty} \frac{\kappa q}{2\pi E_{l,0}(\mu)^2 E_{0,0}(\mu)} \,, \quad K^{(\rm NCP)}_{m,l}(\omega) \approx_{\beta, \omega \rightarrow \infty} \frac{\kappa q}{2\pi E_{0,0}(\mu) \omega^2} \,.
\end{equation}
\end{comment}
Finally, we plug these kernels back into the scaling dimension
\begin{equation}
    \Delta^{(1)}_{(q,0)}(\kappa) =  \int \frac{d \omega}{2\pi} \sum_{l=0}^{\infty} (2l+1) \log \frac{K^{(\rm CP)}_{m,l}(\omega) + K^{(\rm NCP)}_{m,l}(\omega)}{K^{\sigma\sigma}_{0,l}(\omega)} + \mathcal{O}(\kappa^0) \,. 
\end{equation}
Numerically, we verified that the integrand has the following scaling:
\begin{equation}
    \log \frac{K^{(\rm CP)}_{m,l}(\omega) + K^{(\rm NCP)}_{m,l}(\omega)}{K^{\sigma\sigma}_{0,l}(\omega)} = \begin{cases}
        \mathcal{O}(\mu^{1/2}) & l, \omega \lesssim \sqrt{\mu} \\ \mathcal{O}(\mu^0) & l, \omega \gg \sqrt{\mu} 
    \end{cases} \,.
\end{equation}
Therefore, the summation over $l$ and the integral over $\omega$ are dominated by the region where $l, \omega \lesssim \sqrt{\mu}$, which gives a contribution that scales as $\mu^{3/2}$. In the large $\kappa$ limit, the $m$ sector therefore dominates over the $e$ sector and we can estimate the total scaling dimension as
\begin{equation}
    \Delta^{(1)}_{(q,0)}(\kappa) = C_0 \mu^{3/2} + \text{subleading} = C_0 (\kappa q)^{3/2} + \text{subleading} \,,
\end{equation}
where $C_0$ is an undetermined numerical constant. Combining this result with the saddle point answer, we see that
\begin{equation}
    \Delta_{(q,0)}(N, \kappa) \rightarrow_{\kappa \gg 1} \, (\kappa q)^{3/2} \left[\frac{4}{3} N + C_0 + \mathcal{O}(N^{-1}) \right] \,. 
\end{equation}
Generically, $\frac{4}{3}N + C_0 \neq 0$ and quadratic fluctuations preserve the $\kappa^{3/2}$ scaling in the large $\kappa$ limit as promised. 

\subsubsection{Subleading scaling dimension of \texorpdfstring{$\mathcal{M}_{(q,q)}$}{}}\label{app:Mqq_subleading}

The subleading corrections for the balanced monopole $\mathcal{M}_{(q,q)}$ is much simpler because the $e$ and $m$ sectors share a common saddle point solution. As a result, for every $q$, we can write down a single matter field propagator $G_q = \left[- (\nabla - i \bar a)^2 + \frac{1}{4} + \bar \mu\right]^{-1}$ and the matter field contribution to the quadratic action takes the form
\begin{equation}
    S_{2, \phi} = \sum_{\alpha = e, m} S_{2, \alpha} \,, \quad S_{2,\alpha} = \frac{N}{2} \sum_{\alpha} \int \left[a_{\alpha, \mu}(x), \sigma_{\alpha}(x)\right] \begin{pmatrix}
        K_{q}^{\mu\nu}(x,y) & K_{q}^{\mu\sigma}(x,y) \\ K_{q}^{\sigma \nu}(x,y)  & K_{q}^{\sigma\sigma}(x,y) 
    \end{pmatrix} \begin{pmatrix}
        a_{\alpha, \nu}(y) \\ \sigma_{\alpha}(y)
    \end{pmatrix} \,.
\end{equation}
The mutual Chern-Simons term is already quadratic and hence retains its bare form
\begin{equation}
    S_{2, \rm CS} = -\frac{ik}{2\pi} \int a_e d a_m \,. 
\end{equation}
Now observe that the matter kernel is completely independent of $\alpha$, and the mutual Chern-Simons term is proportional to the Pauli-Z matrix. This means that the total quadratic action can be diagonalized in $\alpha$-space via a simple rotation $a_{\pm} = (a_e \pm a_m)/\sqrt{2}, \sigma_{\pm} = (\sigma_e \pm \sigma_m)/\sqrt{2}$. The matter action is left invariant is left invariant while the mutual Chern-Simons term transforms to
\begin{equation}
    S_{2,CS} = - \frac{ik}{2\pi} \int d^3 x \frac{1}{2} (a_+ + a_-) d (a_+ - a_-) = - \frac{ik}{4\pi} \int d^3 x (a_+ d a_+ - a_- d a_-) \,.
\end{equation}
Therefore, the total quadratic action takes the simple form
\begin{equation}
    S_2 = S_{2,+} + S_{2,-} 
\end{equation}
where
\begin{equation}
    S_{2,\pm} = \frac{N}{2} \int \left[a_{\pm, \mu}(x), \sigma_{\pm}(x)\right] \begin{pmatrix}
        K_{q}^{\mu\nu}(x,y) & K_{q}^{\mu\sigma}(x,y) \\ K_{q}^{\sigma \nu}(x,y)  & K_{q}^{\sigma\sigma}(x,y) 
    \end{pmatrix} \begin{pmatrix}
        a_{\pm, \nu}(y) \\ \sigma_{\pm}(y)
    \end{pmatrix} - (\pm) \frac{i k}{4\pi} \int d^3 x a_{\pm} d a_{\pm} \,. 
\end{equation}
This last expression makes it clear that $S_{2,\pm}$ is exactly equal to the quadratic fluctuation action for large $N$ scalar QED3 with self Chern-Simons level $\pm k$. Since the path integral over $+$ and $-$ fields are decoupled, the subleading corrections to the monopole scaling dimension in the self-dual theory are related to the scalar QED3 answer via 
\begin{equation}
    \Delta^{(1)}_{(q,q)}(\kappa) = \Delta^{(1)}_{q, \rm SQED3}(\kappa) + \Delta^{(1)}_{q, \rm SQED3}(-\kappa) = 2\Delta^{(1)}_{q, \rm SQED3}(\kappa) \,.
\end{equation}
Combining this result with the saddle point answer, we reach the simple conclusion that the total scaling dimension is
\begin{equation}
    \begin{aligned}
    \Delta_{(q,q)}(N, \kappa) &= N \Delta^{(0)}_{q,q}(\kappa) + \Delta^{(1)}_{(q,q)}(\kappa) + \mathcal{O}(N^{-1}) \\
    &= 2N \Delta^{(0)}_{q, \rm SQED3}(\kappa) + 2\Delta^{(1)}_{q, \rm SQED3}(\kappa) + \mathcal{O}(N^{-1}) \\
    &= 2 \Delta_{q, \rm SQED3}(N, \kappa) + \mathcal{O}(N^{-1}) \,. 
    \end{aligned}
\end{equation}
This is the simple relation \eqref{eq:Deltaqq_formula} quoted in the main text. From here, we can directly import the results for scalar QED3, which are detailed in Ref.~\cite{C210207377}.

\section{RG analysis of the critical theory}\label{app:RG}

In this appendix, we perform a renormalization group analysis of the critical theory described by \eqref{eq:non_compact_gauge_theory} in the large $N, k$ limit, holding $\kappa = k/N$ fixed. We begin with the large $N$ generalization of the critical Lagrangian:
\begin{equation}
    L = \sum_{\alpha = e, m} \sum_{i=1}^N \left[\left|\left(\nabla - i \frac{a_{\alpha}}{\sqrt{N}}\right) \phi_{i, \alpha}\right|^2 + r |\phi_{i,\alpha}|^2\right] + \frac{v}{N} \sum_{\alpha = e, m} \left(\sum_{i=1}^N |\phi_{i,\alpha}|^2 \right)^2 - \frac{i\kappa}{2\pi} a_e d a_m 
\end{equation}
As is standard in the analysis of large-$N$ vector models, we introduce Hubbard-Stratanovich fields $\sigma_{\alpha}$ to decouple the quartic interactions. After appropriately rescaling the interactions, we can recast the large $N$ effective Lagrangian as a non-linear sigma model
\begin{equation}
    L = \frac{1}{g} \sum_{\alpha=e,m}  \left[\sum_{i=1}^N \left|\left(\nabla - i \frac{a_{\alpha}}{\sqrt{N}}\right) \phi_{i, \alpha}\right|^2 - i v\frac{\sigma_{\alpha}}{\sqrt{N}} \left(\sum_{i=1}^N |\phi_{i,\alpha}|^2 - 1\right)\right] - \frac{i\kappa}{2\pi} a_e d a_m  \,,
\end{equation}
where at the saddle point, $\bar \sigma = - i r$ is a constant that satisfies the UV-regulated equation
\begin{equation}
    \frac{1}{g} = \int \frac{d^d p}{(2\pi)^d} \frac{1}{p^2 + r} \,. 
\end{equation}
The critical point of interest to us is identified with the value $g = g_c$ where $r = 0$. A systematic large $N$ expansion can be performed around this critical point. 

\subsection{Leading order in the large \texorpdfstring{$N$}{} expansion}

\begin{figure}[!h]
    \centering
    \includegraphics[width = 0.8 \textwidth]{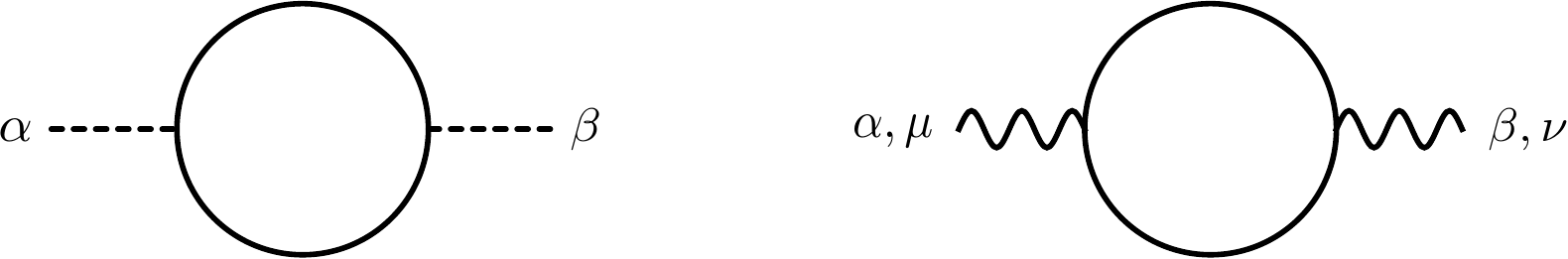}
    \caption{Feynman diagrams that contribute to the self energy of $\sigma_{\alpha}$ and $a_{\alpha}$ at leading order in the large $N$ expansion.}
    \label{fig:leading_order_SE}
\end{figure}

\begin{figure}[!h]
    \centering
    \includegraphics[width=0.9\textwidth]{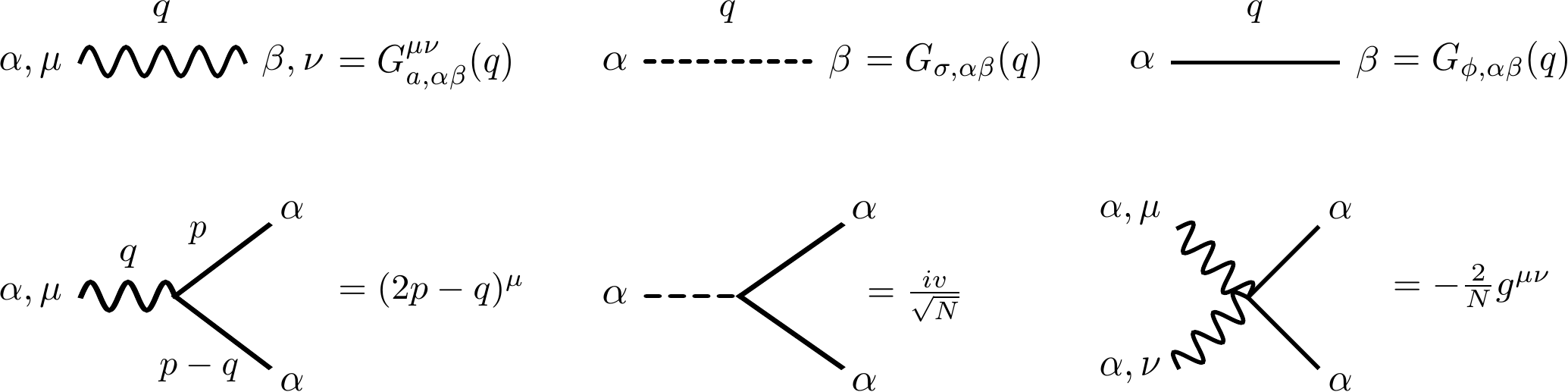}
    \caption{Feynman rules for the critical theory within a generalized Feynman gauge defined in \eqref{eq:feynman_gauge_prop}.}
    \label{fig:feynman_rules}
\end{figure}

At $\mathcal{O}(N^0)$, the only effect of interactions is to generate a singular self-energy for the HS field $\sigma_{\alpha}$ and the gauge fields $a_{\alpha}$. The momentum-independent part of these corrections can be cancelled by a counterterm, and the momentum-dependent parts are captured by the two diagrams in Fig.~\ref{fig:leading_order_SE}.
Using Feynman parameters and dimensional regularization, the self energy for $\sigma_{\alpha}$ and $a^{\mu}_{\alpha}$ can be evaluated:
\begin{equation}    
    \begin{aligned}
    \Pi_{\sigma,\alpha\beta}(q) &= -v^2 \delta_{\alpha\beta} \int \frac{d^d p}{(2\pi)^d} G_{\alpha}(p) G_{\alpha}(p-q) = -v^2 \delta_{\alpha\beta} \int_0^1 dx \int \frac{d^d p}{(2\pi)^d} \frac{1}{\left[x (p-q)^2 + (1-x) p^2\right]^2} \\
    &= -\frac{v^2}{(2\pi)^3} \cdot 4\pi \delta_{\alpha\beta} \int_0^1 dx \lim_{d \rightarrow 3} \int_0^{\infty} \frac{l^{d-1} dl}{\left[l^2 + x (1-x) q^2\right]^2} = -\frac{v^2}{8 q} \delta_{\alpha\beta} \,, \\
    \Pi^{\mu\nu}_{\alpha\beta}(q) &= - \delta_{\alpha\beta} \, \frac{q}{16} \, (g^{\mu\nu} - q^{\mu} q^{\nu}/q^2) \,.
    \end{aligned}
\end{equation}
Since these self-energies are independent of $N$, they must be exactly resummed to generate a new effective Lagrangian
\begin{equation}
    L_{\rm eff} = \frac{1}{g} \sum_{\alpha=e,m}  \left[\sum_{i=1}^N \left|\left(\nabla - i \frac{a_{\alpha}}{\sqrt{N}}\right) \phi_{i, \alpha}\right|^2 - i v\frac{\sigma_{\alpha}}{\sqrt{N}} \left(\sum_{i=1}^N |\phi_{i,\alpha}|^2 - 1\right)\right] + \frac{1}{2} \sigma^{\alpha} \left(G_{\sigma}^{-1}\right)_{\alpha\beta} \sigma^{\beta} + \frac{1}{2} a^{\alpha}_{\mu} (G_a^{-1})^{ \mu\nu}_{\alpha\beta}a^{\beta}_{\nu} \,,
\end{equation}
with
\begin{equation}
    (G_{\sigma}^{-1})_{\alpha\beta}(q) = \frac{v^2}{8 q} \delta_{\alpha\beta} \,, \quad (G_a^{-1})^{\mu\nu}_{\alpha\beta}(q) = \sigma^x_{\alpha\beta} \frac{i\kappa}{2\pi} \epsilon^{\mu\lambda \nu} (i q_{\lambda}) + \frac{q}{16} \delta_{\alpha\beta} (g^{\mu\nu} - q^{\mu} q^{\nu}/q^2) \,.  
\end{equation}
Before proceeding further, we need to fix a gauge to make the gauge field propagator $G_a$ well-defined. Following the Fadeev-Popov procedure, we define a family of gauges parametrized by $\xi$ so that the kernel gets deformed to
\begin{equation}
    (G_a^{-1})^{\mu\nu}_{\alpha\beta}(\xi, q) = \sigma^x_{\alpha\beta} \frac{i\kappa}{2\pi} \epsilon^{\mu\lambda \nu} (i q_{\lambda}) + \frac{q}{16} \delta_{\alpha\beta} \left[g^{\mu\nu} - (1 - \xi^{-1})q^{\mu} q^{\nu}/q^2\right] \,. 
\end{equation}
In the rotated basis $a_{\pm} = (a_e \pm a_m)/\sqrt{2}$, the kernel becomes diagonal
\begin{equation}
    (G_a^{-1})^{\mu\nu}_{\pm}(\xi, q) = \frac{\pm i\kappa}{2\pi} L^{\mu\nu} + \frac{1}{16q} \left[(L^2)^{\mu\nu} + \xi^{-1} q^{\mu} q^{\nu} \right] \,, \quad L^{\mu\nu} = \epsilon^{\mu\lambda\nu} (i q_{\lambda}) \,.
\end{equation}
A general kernel of the form $A L^{\mu\nu} + B q^{-1} \left[(L^2)^{\mu\nu} + \xi^{-1} q^{\mu} q^{\nu} \right]$ can be inverted by the ansatz
\begin{equation}
    G^{\mu\nu}_{a,\pm}(\xi, q) = C L^{\mu\nu} + D q^{-1} \left[(L^2)^{\mu\nu} + f(\xi) q^{\mu} q^{\nu} \right] \,, \quad  AD + BC = AC + BD - BD \xi^{-1} f(\xi) = 0 \,, \quad q^2 (AC + BD) = 1 \,. 
\end{equation}
The explicit solution for arbitrary $A, B$ is given by
\begin{equation}
    C = \frac{A q^{-2}}{A^2 - B^2} \,, \quad D = \frac{-B q^{-2}}{A^2 - B^2} \,, \quad f(\xi) = \xi \left[1 - \frac{A^2}{B^2}\right] \,.
\end{equation}
For calculational convenience, we will choose the generalized Feynman gauge $\xi_* = \left[1 - \frac{A^2}{B^2}\right]^{-1}$ so that $f(\xi_*) = 1$. After plugging in $A = \pm \frac{i\kappa}{2\pi}, B = 1/16$, we find the gauge propagator
\begin{equation}\label{eq:feynman_gauge_prop}
    G^{\mu\nu}_{a, \pm}(\xi_*, q) = \pm C(\kappa) \frac{L^{\mu\nu}}{q^2} + D(\kappa) \frac{g^{\mu\nu}}{q} \,, \quad C(\kappa) = - \frac{128 i \kappa}{\pi\left[1 + \frac{64 \kappa^2}{\pi^2}\right]} \,, \quad D(\kappa) = \frac{16}{1 + \frac{64 \kappa^2}{\pi^2}} \,. 
\end{equation}
Rotating back to the original basis and dropping the $\xi_*$ label, we obtain the propagators
\begin{equation}
    \begin{aligned}
    G^{\mu\nu}_{a, \alpha\alpha}(q) &= \frac{1}{2} \left[ G^{\mu\nu}_{a, +}(q) + G^{\mu\nu}_{a, -}(q)\right] = D(\kappa) \frac{g_{\mu\nu}}{q} \,, \\
    G^{\mu\nu}_{a, em}(q) &= \frac{1}{2} \left[ G^{\mu\nu}_{a, +}(q) - G^{\mu\nu}_{a, -}(q)\right] = C(\kappa) \frac{L^{\mu\nu}}{q^2} \,. 
    \end{aligned}
\end{equation}
With these explicit gauge-fixed propagators, we can infer the Feynman rules in Fig.~\ref{fig:feynman_rules} and compute $1/N$ corrections.

\subsection{Subleading order in the large \texorpdfstring{$N$}{} expansion: renormalization of \texorpdfstring{$\sigma_{\alpha}$}{} (equivalently \texorpdfstring{$|\phi_{\alpha}|^2$}{})}

\begin{figure}[!h]
    \centering
    \includegraphics[width = 0.9\textwidth]{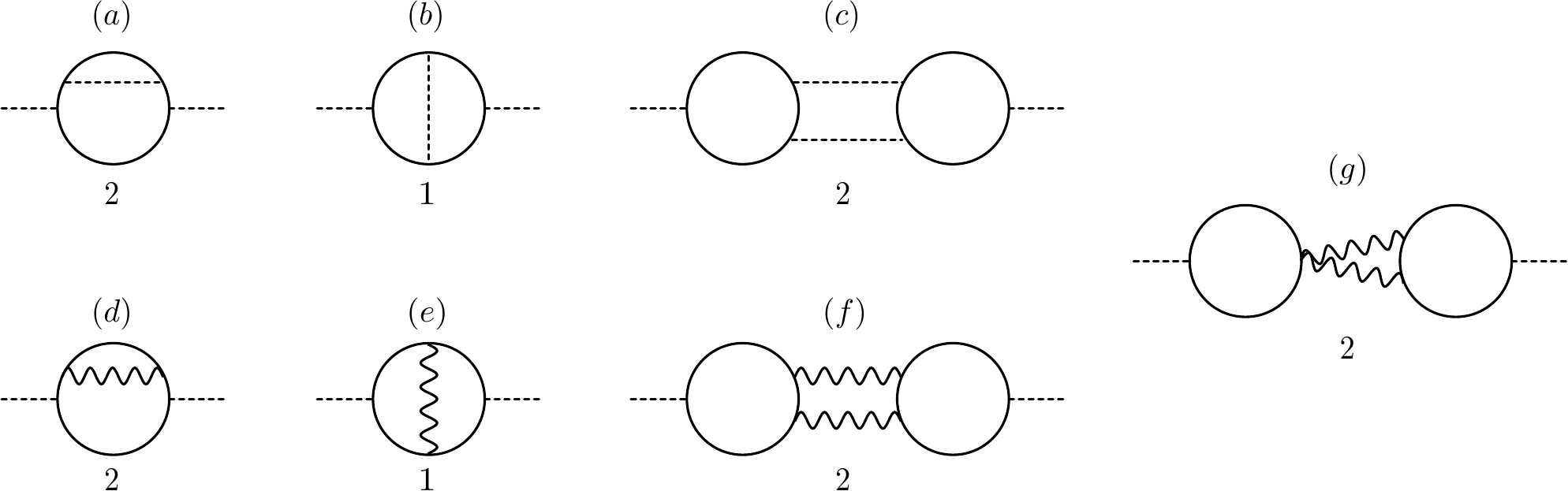}
    \caption{Diagrams contributing to the self energy $\Pi_{\sigma, \alpha\beta}$. The number below each diagram indicates the symmetry factors.}
    \label{fig:subleading_order_sigma_SE}
\end{figure}

We first consider the renormalization of $\sigma_{\alpha}$. At $\mathcal{O}(N^{-1})$, the total self-energy $\Pi_{\sigma, \alpha\beta}(q)$ can be decomposed into
\begin{equation}
    \Pi_{\sigma, \alpha\beta}(q) = \sum_i \Pi^{(i)}_{\sigma, \alpha\beta}(q) \,,
\end{equation}
where $i$ labels all the diagrams that depend on external momentum, as shown in Fig.~\ref{fig:subleading_order_sigma_SE}.
By power counting, we see that the integrands in diagrams (c) and (f) scale as $p^{-10}$, while the total number of momentum integrals is $3d$. For $d = 3$, since there is no divergent subdiagram, the integral is UV-convergent and cannot generate logarithmic singularities. Moreover, the index structure of the interaction vertices in Fig.~\ref{fig:feynman_rules} imply that diagrams (a), (b), (c), (d), (e) can only generate singularities proportional to $\delta_{\alpha\beta}$, while diagram (g) can generate a diagonal term as well as an off-diagonal term proportional to $\sigma^x_{\alpha\beta}$. 

We first evaluate the diagonal terms. We will use dimensional regularization at $d = 3 - \epsilon$ and trade each factor of $1/\epsilon$ with $- \log q$ where $q$ is the external momentum. For example, diagram (a) can be evaluated as 
\begin{equation}
    \begin{aligned}
        \Pi^{(a)}_{\sigma,\alpha\alpha}(q) &= 2 \cdot \frac{v^4}{N} \int \frac{d^d p}{(2\pi)^d} \int \frac{d^d q_1}{(2\pi)^d}  G_{\phi,\alpha\alpha}(p)^2 G_{\phi,\alpha\alpha}(p-q) G_{\phi,\alpha\alpha}(p-q_1) G_{\sigma,\alpha\alpha}(q_1) \\
        &= \frac{2v^4}{N} \int \frac{d^d p}{(2\pi)^d} \frac{1}{p^4 (p-q)^2} \int \frac{d^d q_1}{(2\pi)^d} \frac{8}{v^2 q_1} \left[1 + \frac{2 q_1 \cdot p - p^2}{q_1^2} + \frac{(2 q_1 \cdot p - p^2)^2}{q_1^4} + \ldots \right] \\
        &= \frac{16v^2}{N} \int \frac{d^d p}{(2\pi)^d} \frac{1}{p^2 (p-q)^2} \frac{1}{4\pi^2} \int_0^{\pi} d \theta \sin \theta \int_0^{\infty}  dq_1 \, q_1^{d-4} (4 \cos^2 \theta - 1) + \text{non-log} \\
        &= \frac{16 v^2}{N} \cdot \frac{1}{8q} \cdot \frac{1}{4\pi^2} \cdot \frac{2}{3 \epsilon} = - \frac{v^2}{8 q} \cdot \frac{8}{3\pi^2 N} \log q \,.
    \end{aligned}
\end{equation}
Using a similar trick and replacing $\sigma$-propagators with gauge propagators, we find that
\begin{equation}
    \Pi^{(d)}_{\sigma,\alpha\alpha}(q) = - \frac{v^2}{8q} \cdot - \frac{5}{3\pi^2 N} D(\kappa) \log q \,.
\end{equation}
Next, we turn to the vertex corrections in diagrams (b) and (e). For (b), it is convenient to write the loop integral in a symmetric fashion:
\begin{equation}    
    \begin{aligned}
    \Pi^{(b)}_{\sigma,\alpha\alpha}(q) &= \frac{v^4}{N} \int \frac{d^d p_1}{(2\pi)^d} \frac{d^d p_2}{(2\pi)^d} G_{\phi,\alpha\alpha}(p_1) G_{\phi,\alpha\alpha}(p_1-q) G_{\phi,\alpha\alpha}(p_2) G_{\phi,\alpha\alpha}(p_2 - q) G_{\sigma,\alpha\alpha}(p_1-p_2) \\
    &= \frac{8v^2}{N} \int \frac{d^d p_1}{(2\pi)^d} \frac{d^d p_2}{(2\pi)^d} \frac{|p_1-p_2|}{p_1^2 (p_1-q)^2 p_2^2 (p_2-q)^2} \,.
    \end{aligned}
\end{equation}
The integral is superficially convergent in the integration region where $p_1, p_2 \rightarrow \infty$ with $p_1/p_2$ fixed. However, there are divergences coming from the integration region where $|p_2| \ll |p_1|$ or $|p_1| \ll |p_2|$. These divergences can be interpreted as vertex corrections dressing the left/right vertex in diagram (b). Due to the symmetry between $p_1, p_2$, these divergences are identical and we simply need to compute the integral in one region and multiply by 2. Without loss of generality, let us work in the regime $|p_2| \ll |p_1|$:
\begin{equation}
    \begin{aligned}
        \Pi^{(b)}_{\sigma,\alpha\alpha}(q) &\approx  2 \cdot \frac{8 v^2}{N} \int \frac{d^d p_1}{(2\pi)^d} \frac{d^d p_2}{(2\pi)^d} \frac{p_1}{p_1^2(p_1-q)^2} \frac{1}{p_2^2 (p_2-q)^2} = \frac{16 v^2}{N} \frac{1}{8q} \int \frac{d^d p_1}{(2\pi)^d} \frac{p_1}{p_1^2(p_1-q)^2} + \text{non-log} \\
        &\approx \frac{2 v^2}{N q} \cdot \frac{4 \pi}{8\pi^3} \cdot \int p_1^{d-4} d p_1 = \frac{v^2}{\pi^2 N q \epsilon} = - \frac{v^2}{8q} \cdot \frac{8}{\pi^2 N} \log q \,. 
    \end{aligned}
\end{equation}
Using a similar trick, one can easily calculate diagram (e)
\begin{equation}
    \Pi^{(e)}_{\sigma,\alpha\alpha} = - \frac{v^2}{8q} \cdot \frac{-1}{\pi^2 N} D(\kappa) \log q  \,.
\end{equation}
Finally, we come to diagram (g). We first consider the diagonal case, where the integral can be represented as
\begin{equation}
    \begin{aligned}
        \Pi^{(g)}_{\sigma,\alpha\alpha}(q) &= 2 \cdot \frac{-v^2}{N} \int \frac{d^d p_1}{(2\pi)^d} G_{\phi,\alpha\alpha}(p_1) G_{\phi,\alpha\alpha}(p_1-q) \cdot (-2) g_{\mu\nu} \int \frac{d^d q_1}{(2\pi)^d} G^{\mu \mu_1}_{a,\alpha\alpha}(q_1) G^{\nu\nu_1}_{a, \alpha\alpha}(q_1-q) \\
        &\int \frac{d^d p}{(2\pi)^d} (2p - q_1)^{\mu_1} (2p - q - q_1)^{\nu_1} G_{\phi,\alpha\alpha}(p) G_{\phi,\alpha\alpha}(p-q) G_{\phi,\alpha\alpha}(p-q_1) \\
        &= \frac{v^2}{2Nq} D(\kappa)^2 g_{\mu\nu} \int \frac{d^d q_1}{(2\pi)^d} \int \frac{d^d p}{(2\pi)^d} \frac{g^{\mu\mu_1} g^{\nu\nu_1}}{q_1^2} \frac{(2p - q_1)^{\mu_1} (2p - q_1)^{\nu_1}}{p^4 (p-q_1)^2} + \text{non-log} \\
        &= \frac{D(\kappa)^2 v^2}{2Nq} \int \frac{d^d p}{(2\pi)^d} \frac{1}{p^4} I(p) + \text{non-log} 
    \end{aligned}
\end{equation}
with $I(p)$ representing a one-loop integral:
\begin{equation}
    I(p) = \int \frac{d^d q_1}{(2\pi)^d} \frac{(q_1-2p)^2}{q_1^2 (q_1-p)^2} \,. 
\end{equation}
This one-loop integral can be done by Feynman parameters:
\begin{equation}
    \begin{aligned}
    I(p) &= \int_0^1 dx \int \frac{d^d l}{(2\pi)^d} \frac{(l + xp - 2p)^2}{\left[l^2 + x(1-x) p^2\right]^2} = \frac{1}{4\pi^2} \int_0^1 dx \int_0^{\pi} d \theta \sin \theta \int_0^{\infty} dl  \frac{l^{d-1}\left[l^2 + 2(x-2) l p \cos \theta + (x-2)^2 p^2\right] }{\left[l^2 + x(1-x) p^2\right]^2} \\
    &= \frac{1}{2\pi^2} \int_0^1 dx \int_0^{\infty} dl \frac{l^{d-1}}{\left[l^2 + x(1-x) p^2\right]^2} \left[l^2 + (x-2)^2 p^2\right] = \frac{p}{2\pi^2} \int_0^1 dx \left[- \frac{3 \pi \sqrt{x(1-x)}}{4} + \frac{\pi (x-2)^2}{4\sqrt{x(1-x)}}\right] = \frac{p}{4} \,. 
    \end{aligned}
\end{equation}
Plugging this answer back into the integral over $p$, we find 
\begin{equation}
    \Pi^{(g)}_{\sigma,\alpha\alpha}(q) = \frac{D(\kappa)^2 v^2}{2Nq} \frac{1}{2\pi^2} \int d p \, \frac{p^{d-1}}{p^4} \frac{p}{4}  = \frac{D(\kappa)^2 v^2}{16 \pi^2 Nq \epsilon} = - \frac{v^2}{8q} \cdot \frac{D(\kappa)^2}{2 \pi^2 N} \log q \,. 
\end{equation}
Next, we come to the off-diagonal contribution. After plugging in the off-diagonal gauge field propagators in \eqref{eq:feynman_gauge_prop}, we can simplify the integral to 
\begin{equation}
    \begin{aligned}
    \Pi^{(g)}_{\sigma,em}(q) &= 2 \cdot \frac{-v^2}{N} \int \frac{d^d p_1}{(2\pi)^d} G_{\phi,ee}(p_1) G_{\phi,ee}(p_1-q) \cdot (-2) g_{\mu\nu} \int \frac{d^d q_1}{(2\pi)^d} G^{\nu_1\nu}_{a,me}(q_1-q) G^{\mu \mu_1}_{a,em}(q_1)  \\
    &\int \frac{d^d p}{(2\pi)^d} (2p - q_1)_{\mu_1} (2p - q - q_1)_{\nu_1} G_{\phi, mm}(p) G_{\phi,mm}(p-q) G_{\phi,mm}(p-q_1) \\
    &\approx \frac{v^2}{2Nq} g_{\mu\nu} \int \frac{d^d p}{(2\pi)^d} \frac{1}{p^4} \int \frac{d^d q_1}{(2\pi)^d} C(\kappa)^2 \frac{L^{\nu_1\nu} L^{\mu\mu_1}}{q_1^4} \frac{(q_1 - 2p)_{\mu_1} (q_1 - 2p)_{\nu_1}}{(p-q_1)^2} \\
    &= \frac{v^2 C(\kappa)^2}{2Nq} \int \frac{d^d p}{(2\pi)^d} \frac{1}{p^4} J(p) \,,
    \end{aligned}
\end{equation}
where
\begin{equation}
    J(p) = \int \frac{d^d q_1}{(2\pi)^d} \frac{q_1^2 g^{\mu_1 \nu_1} - q_1^{\mu_1} q_1^{\nu_1}}{q_1^4} \frac{(q_1 - 2p)_{\mu_1} (q_1 - 2p)_{\nu_1}}{(p-q_1)^2} \,.
\end{equation}
To do this integral, let us invoke the generalized Feynman parameters trick
\begin{equation}
    \frac{1}{A^m B^n} = \int_0^{\infty} \frac{\lambda^{m-1} d \lambda}{\left[\lambda A + B\right]^{m+n}} \,. 
\end{equation}
Applying this trick with $A = q_1^2, B = (q_1-p)^2, m = 2, n = 1, d = 3$, we obtain
\begin{equation}
    \begin{aligned}
        J(p) &= 4 \int \frac{d^3 q_1}{(2\pi)^3} \frac{q_1^2 p^2 - (q_1 \cdot p)^2}{q_1^4 (q_1 - p)^2} = 4 p_{\mu_1} p_{\nu_1} \int_0^{\infty} \lambda d \lambda \int \frac{d^3 q_1}{(2\pi)^3} \frac{q_1^2 g^{\mu_1\nu_1} - q_1^{\mu_1} q_1^{\nu_1}}{\left[\lambda q_1^2 + (q_1 - p)^2\right]^3} \\
        &= 4 p_{\mu_1} p_{\nu_1} \int_0^{\infty} \lambda d \lambda \int \frac{d^3 l_1}{(2\pi)^3} \frac{\left(l_1 + \frac{p}{\lambda+1}\right)^2 g^{\mu_1\nu_1} - \left(l_1 + \frac{p}{\lambda+1}\right)^{\mu_1} \left(l_1 + \frac{p}{\lambda+1}\right)^{\nu_1}}{\left[(\lambda+1) l_1^2 + \frac{\lambda}{\lambda+1} p^2\right]^3}  \\
        &= \frac{8p^2}{3} \int_0^{\infty} \lambda d \lambda \frac{1}{2\pi^2} \int_0^{\infty} \frac{l_1^4}{\left[(\lambda+1) l_1^2 + \frac{\lambda}{\lambda+1} p^2\right]^3} = \frac{4 p^2}{3 \pi^2} \int_0^{\infty} d \lambda \frac{3 \pi \lambda}{16 p \sqrt{\lambda}(\lambda+1)^2} = \frac{p}{8} \,. 
    \end{aligned}
\end{equation}
Plugging $J(p)$ back into $\Pi^{(g)}_{em}(q)$, we find that
\begin{equation}
    \Pi^{(g)}_{\sigma,em}(q) = \frac{v^2 C(\kappa)^2}{2Nq} \frac{1}{2\pi^2} \int \frac{dp}{p^2} \frac{p}{8} = - \frac{v^2}{8q} \frac{C(\kappa)^2}{4\pi^2 N} \log q \,.
\end{equation}
Putting all the terms together, we obtain the final answer
\begin{equation}
    \begin{aligned}
    \Pi_{\sigma,ee/mm}(q) &= - \frac{v^2}{8q} \log q \left(\frac{8}{3\pi^2N} - \frac{5 D(\kappa)}{3\pi^2N} + \frac{8}{\pi^2 N} - \frac{D(\kappa)}{\pi^2N} + \frac{D(\kappa)^2}{2\pi^2N} \right) = - \frac{v^2}{8q} \log q \cdot \frac{16 + 48 - 16 D(\kappa) + 3 D(\kappa)^2}{6 \pi^2 N} \,,\\
    \Pi_{\sigma,em/me}(q) &= - \frac{v^2}{8q} \log q \cdot  \frac{C(\kappa)^2}{4\pi^2 N}  \,.
    \end{aligned}
\end{equation}
After diagonalizing this matrix and plugging in the explicit expressions for $C(\kappa)$ and $D(\kappa)$, we find that
\begin{equation}
    \begin{aligned}
    \Pi_{\sigma,\pm \pm}(q) &= - \frac{v^2}{8q} \log q \left[\frac{32}{3\pi^2N} - \frac{128}{3\pi^2N \left[1 + \frac{64 \kappa^2}{\pi^2}\right]} + \frac{128}{\pi^2 N \left[1 + \frac{64 \kappa^2}{\pi^2}\right]^2} \mp \frac{64^2 \kappa^2}{\pi^4 N \left[1 + \frac{64 \kappa^2}{\pi^2}\right]^2} \right] \,.
    \end{aligned}
\end{equation}
In an RG interpretation, anomalous dimensions for $\sigma_{\pm}$ enter the self energies via 
\begin{equation}
    \Pi_{\sigma, \pm \pm}(q) \sim -\frac{v^2}{8q} q^{-2 \eta_{\sigma_{\pm}}} \sim -\frac{v^2}{8q} (1 - 2 \eta_{\sigma_{\pm}} \log q + \ldots ) \,. 
\end{equation}
Using this definition, we can extract the anomalous dimensions 
\begin{equation}
    \eta_{\sigma_{\pm}}(\kappa) = - \frac{16}{3\pi^2N} + \frac{64}{3\pi^2 N \left[1 + \frac{64 \kappa^2}{\pi^2}\right]} - \frac{64}{\pi^2 N \left[1 + \frac{64 \kappa^2}{\pi^2}\right]^2} \pm \frac{64^2 \kappa^2}{2 \pi^4 N \left[1 + \frac{64 \kappa^2}{\pi^2}\right]^2} \,. 
\end{equation}
From here, the correlation length exponent immediately follows:
\begin{equation}
    \nu^{-1} = d - \Delta_{\sigma_+} = 1 - \eta_{\sigma_+} \,. 
\end{equation}
As a sanity check, note that when $\kappa = 0$, the $e,m$ sectors completely decouple and we recover the $\mathcal{O}(N^{-1})$ anomalous dimensions in the $\mathbb{CP}^{N-1}$ model
\begin{equation}
    \eta_{\sigma_{\pm}}(\kappa) = - \frac{48}{\pi^2 N} \,. 
\end{equation}
In the opposite limit, we see that all corrections to the scaling dimension vanish as $\kappa \rightarrow \infty$
\begin{equation}
    \eta_{\sigma_{\pm}}(\kappa) \approx - \frac{16}{3\pi^2N} + \frac{1}{3 N \kappa^2} \pm \frac{1}{2N \kappa^2} + \mathcal{O}(\kappa^{-4}) \,. 
\end{equation}
This result is in agreement with our general argument (valid for all $N$) that anomalous dimensions are suppressed by a factor of $k^{-2}$ in the large $k$ limit.

\subsection{Subleading order in the large \texorpdfstring{$N$}{} expansion: renormalization of the Chern-Simons level \texorpdfstring{$k$}{}}\label{app:CS_level_RG}

In the previous calculation, the Chern-Simons level $k$ is held fixed throughout the calculation. This is justified because the presence of monopoles in the effective Lagrangian guarantees that the Chern-Simons level is quantized, even though the monopoles are irrelevant in the RG sense. Here, we verify that this is indeed the case at $\mathcal{O}(N^{-1})$. 

To see that, we can draw all 1PI self energy diagrams for the gauge fields and label their contributions as $\Pi^{(i),\mu\nu}_{\alpha\beta}$. The non-vanishing diagrams have internal structures that are identical to the ones in Fig.~\ref{fig:subleading_order_sigma_SE}, except the external scalar lines are replaced by gauge field lines. To renormalize the Chern-Simons level, we need to find a diagram that mixes $e, m$ sectors. The only such diagram is given by $(g)$ with external lines replaced by gauge field lines. However, this diagram contains a subdiagram which is the three-point function of the current operator in a free scalar field theory, which vanishes identically. Therefore, there is no RG flow for $k$ at $\mathcal{O}(N^{-1})$. %It is an interesting future direction to search for the leading non-vanishing diagram at higher order that renormalizes $k$. 

\subsection{Subleading order in the large \texorpdfstring{$N$}{} expansion: renormalization of the scalar field}

Finally, we include here the $\mathcal{O}(N^{-1})$ anomalous dimensions for the scalar field $\phi_{\alpha}$. Although these anomalous dimensions are gauge-dependent, they are useful as intermediate steps in the calculation of anomalous dimensions for gauge-invariant operators such as the quartic operators $|\phi_{\alpha}|^4$ and $|\phi_e|^2 |\phi_m|^2$. We leave these evaluations to future work. 

\begin{figure}[!h]
    \centering
    \includegraphics[width = 0.7 \textwidth]{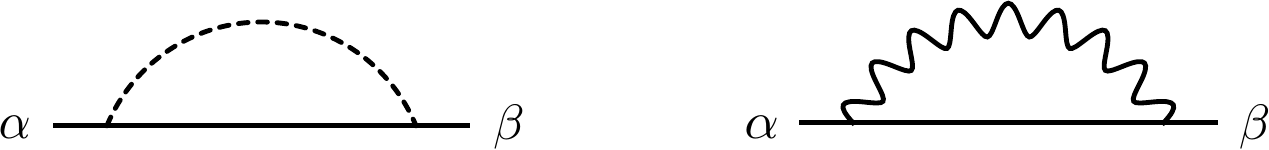}
    \caption{Diagrams that contribute to the non-gauge-invariant renormalization of $\phi_{\alpha}$.}
    \label{fig:subleading_order_phi_SE}
\end{figure}
As shown in Fig.~\ref{fig:subleading_order_phi_SE}, there are only two momentum-dependent corrections to the scalar field self energy $\Pi_{\phi, \alpha\beta}$. The total contribution of these diagrams is given by
\begin{equation}
    \eta_{\phi_{\alpha}} = \frac{4}{3 \pi^2 N} - \frac{5}{6 \pi^2 N} D(\kappa) = \frac{4}{3\pi^2 N} - \frac{40}{3 \pi^2 N \left[1 + \frac{64 \kappa^2}{\pi^2}\right]} \,.
\end{equation}
Combining this result with the result for $\nu^{-1}$, we can also infer the exponent $\gamma$ that controls the divergence of the boson susceptibility as the critical point is approached
\begin{equation}
    \gamma = \nu (2 - \eta_{\phi_{\alpha}}) = \frac{2 - \eta_{\phi_{\alpha}}}{1 - \eta_{\sigma_+}} \,. 
\end{equation}

%\nocite{*}

\end{document}